\documentclass[10pt,a4paper]{article}

\usepackage{graphicx,latexsym}
\graphicspath{{images/}{../images/}}
\usepackage{amssymb,amsthm,amsmath,amsfonts,bm}
\usepackage{dcolumn}
\usepackage{gensymb}
\usepackage{braket}
\usepackage{siunitx}
\usepackage{layouts}
\usepackage{lipsum}
\usepackage{float}
\usepackage{caption}
\usepackage{nameref}
\usepackage{cleveref}

\usepackage[margin=1in]{geometry}
\renewcommand{\thefigure}{\arabic{figure}}
\usepackage{comment}

\newcommand{\AffOxEng}{Department of Engineering Science, University of Oxford, Parks Road, Oxford OX1 3PJ, UK}
\newcommand{\AffOxMat}{Department of Materials, University of Oxford, Parks Road, Oxford OX1 3PH, UK}

\begin{document}

\title{Raman Microspectroscopy for Real-Time Structure Indicator in Ultrafast Laser Writing}

\author{
    Xingrui Cheng$^{1,*}$,
    Eugenio Picheo$^{2}$,
    Zhixin Chen$^{1,2}$,\\
    Martin J.\ Booth$^{1}$,
    Patrick S.\ Salter$^{1}$,
    Álvaro Fernández\textendash Galiana$^{1,*}$
}

\date{}
\maketitle

\begin{center}
\small
$^{1}$ \AffOxEng \\
$^{2}$ \AffOxMat \\
\vspace{4pt}
$^{*}$Corresponding authors: \texttt{xingrui.cheng@eng.ox.ac.uk},
\texttt{alvaro.fernandezgaliana@eng.ox.ac.uk}
\end{center}

\begin{abstract}
Femtosecond laser fabrication enables the creation of a wide range of devices, but its scalability and yield can be limited by the lack of real-time, in-situ monitoring tools. In particular, there is a strong need for metrics that directly correlate with device performance. Raman microspectroscopy provides a non-destructive route for in-situ characterization. Here, we demonstrate its potential to assess the electrical performance of laser-written graphitic electrodes in diamond. By combining hyperspectral mapping with electrical testing, we show that depletion of the 1332~cm$^{-1}$ sp3 Raman line serves as a monotonic and robust predictor of resistance, offering clear advantages over commonly used spectral features. We further introduce hyperspectral unmixing as a label-free approach to identify relevant spectral signatures in fabrication processes where Raman markers are less defined. Importantly, the methodology we present is not restricted to diamond but can be adapted to other host materials and functionalities, offering a practical path toward specification-driven fs-laser microfabrication.
\end{abstract}



\maketitle

\section{Introduction}
\label{sec:Intro}

Femtosecond laser writing enables direct, three-dimensional patterning of embedded features in transparent hosts via nonlinear absorption—multi-phonon ionization and Zener breakdown—that initiates localized structural transformation~\cite{gattass2008femtosecond, musgraves2011laser}.
The technique has found particular application in processing of diamond~\cite{britel2025comparative, Jeschke1999microscopic,Ashikkalieva2016,Kononenko2016,sotillo2016},  permitting fabrication of buried and surface carbon wires and electrode architectures with arbitrary geometry for electrically conductive devices~\cite{Sun2014}, such as radiation detectors~\cite{komlenok2016diamond, Lagomarsino2013}. 
Yet, one of the main current limitations for scaling manufacturing of these laser-written devices is the lack of feedback mechanisms that can be integrated for real-time, in-situ process monitoring and control. Indeed, most current structural analyses remain destructive or ex-situ (e.g.,TEM~\cite{Ashikkalieva2019,Salter2024}, SEM~\cite{Ashikkalieva2016}, and X-ray diffraction microscopy~\cite{Yin2024}). 

Raman microspectroscopy, based on inelastic light scattering, has been extensively used for non-destructive identification, structural characterization, and monitoring of chemical and physical properties across materials science, chemistry, and biology~\cite{fernandez2024fundamentals}. 
In recent years, significant efforts have also focused on developing advanced computational approaches for spectral analysis, further enhancing and broadening its range of applications. 

Given its ability to differentiate the spectral signatures of sp3 versus sp2 bonding, crystallinity, and local strain environment, Raman spectroscopy is ideal for non-invasive, in-situ, spatially resolved characterization of carbon systems~\cite{khomich2022raman,Sotillo2018}. 
In carbon systems, Raman signatures provide direct markers of bonding and structural order: graphitic sp2 domains give rise to the G band near $\sim1580~\mathrm{cm^{-1}}$, while disorder or finite crystallite size activates the D band at $\sim1350{-}1360~\mathrm{cm^{-1}}$~\cite{Tuinstra1970,Ferrari2000}. In contrast, single-crystal diamond exhibits a sharp first-order mode at $1332~\mathrm{cm^{-1}}$~\cite{Knight1989,Ferrari2004}. Throughout this work, we denote these features as the sp3 line ($1332~\mathrm{cm^{-1}}$), sp2 band ($\sim1580~\mathrm{cm^{-1}}$), and D band ($\sim1350{-}1360~\mathrm{cm^{-1}}$), providing a consistent basis for assessing carbon phase composition, crystallinity, and defect populations in laser-written structures.

Here, we use Raman microspectroscopy to systematically assess fs-laser-written graphitic electrodes in diamond, and compare its performance with brightfield and photoluminescence imaging. Leveraging phase-specific vibrational modes of carbon, we perform direct mapping of surface graphitization through explicit band assignment and label-free spectral deconvolution, and correlate the spectral signatures to the device performance (i.e., electrical conductivity). Beyond offering detailed insights into graphitization monitoring in diamond, the methodology outlined here is broadly transferable: it can be applied to evaluate Raman spectroscopy as a performance probe in diverse device platforms, since none of its principles are inherently diamond-specific.

\section{Results and Discussion}
\label{sec:Experimental Overview}

\begin{figure}[H]
    \centering
    \includegraphics[width=\linewidth]{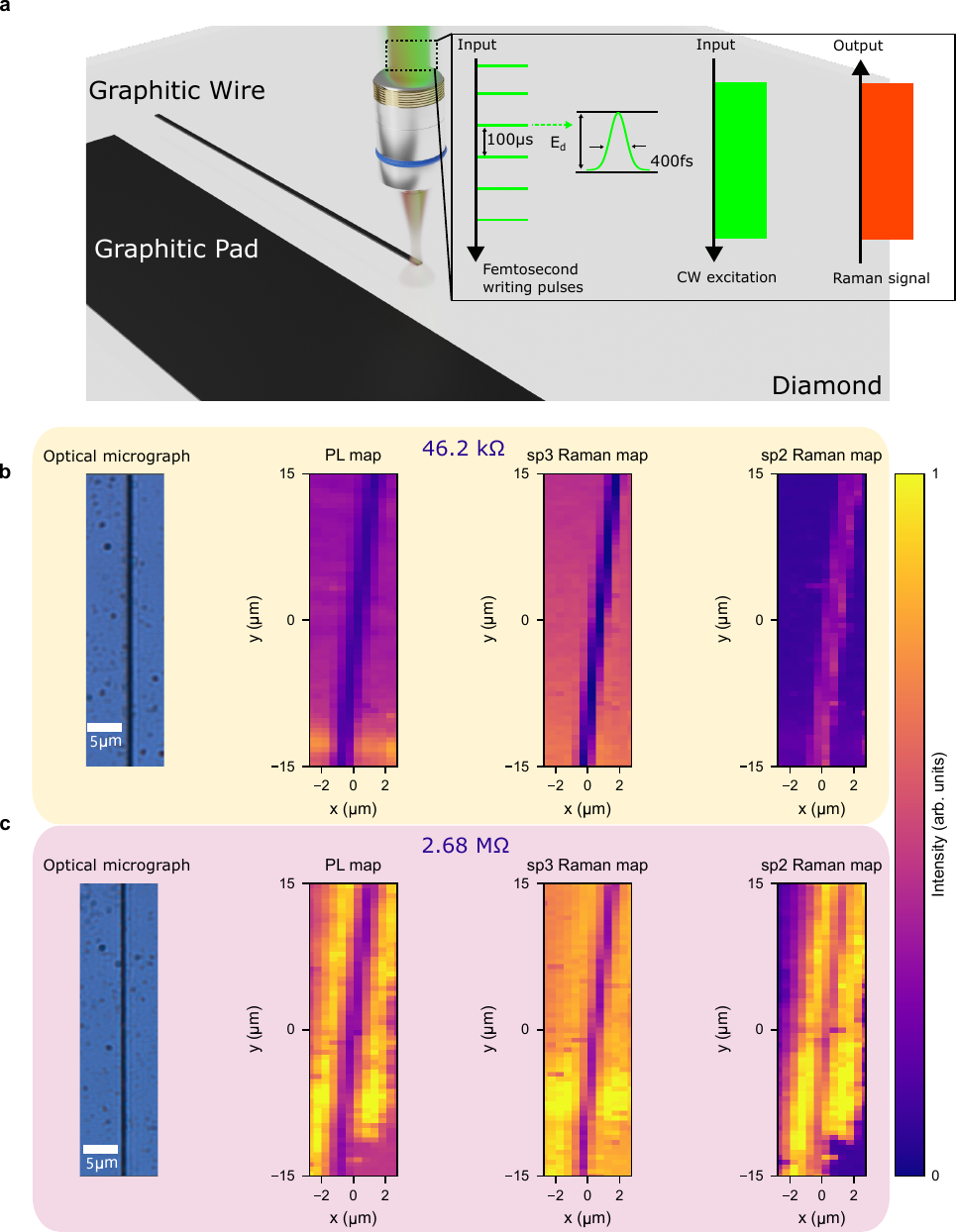}
    \caption{\textbf{Laser-written graphitic structures with optical/PL/Raman characterization.}~\textbf{a)} Schematic of fs-laser fabrication of graphitic pads and wires. The fabrication beam shares a common optical path with the 532~nm Raman excitation, and the back-scattered Raman signal is collected to enable in situ monitoring and post-fabrication mapping.
    \textbf{b-c)} Optical micrographs, PL maps, and Raman intensity maps (sp3 and sp2 windows) for wires written at scan speeds of 20~$\mu$m~s$^{-1}$ and 200~$\mu$m~s$^{-1}$, respectively, collected after fabrication. For direct comparison, PL and Raman maps are unprocessed and share identical count limits. The 20~$\mu$m~s$^{-1}$ wire exhibits a resistance of $46.2~\mathrm{k}\Omega \pm 41.2~\Omega$, whereas the 200~$\mu$m~s$^{-1}$ wire exhibits $2.68~\mathrm{M}\Omega \pm 836~\Omega$. The stronger optical darkening at the lower scan speed provides a first qualitative indication of increased graphitization. However, the contrast in optical micrographs and PL maps alone is difficult to resolve quantitatively. By contrast, Raman mapping of the sp3 window provides a direct and more reliable measure of the degree of graphitization, offering phase-specific insight that complements the broadband PL response and sp2 contrast.}
\label{fig:1}
\end{figure}

To model the principal conductive elements in diamond devices, we use a femtosecond laser to fabricate graphitic pads and wires on the surface of a CVD diamond sample, as shown in Figure~\ref{fig:1}a. Pads wider than 25~$\mu$m and 200~$\mu$m long reproduce the large‐area contact zones required for wire bonding or subsequent metallization, where sheet resistance and adhesion set the performance limit. By contrast, 1~$\mu$m $\times$ 200~$\mu$m wires replicate the sub-micron buried interconnects already deployed in high-density pixel detectors, biosensing micro-electrodes, and microwave striplines for NV-center control~\cite{Salter2024,Bloomer2020,FORNERIS2017}. Studying both geometries spans the full electrical landscape of laser-modified diamond.
Moreover, pads and wires of equal length were fabricated at varying laser scanning speeds to produce different degrees of graphitization, and thus a controlled range of electrical conductivities. 
For each of these structures, brightfield transmission microscopy images were captured, along with photoluminescence (PL) and backscattered Raman, collected using our integrated confocal microscope (see \nameref{sec:Method}).

Figures~\ref{fig:1}b and c show these optical micrographs along with PL and Raman maps (sp3 and sp2) for two laser-written graphitic wires fabricated at scan speeds of 20 and 200~$\mu$m~s$^{-1}$. The wire written at 20~$\mu$m~s$^{-1}$ exhibits a resistance of $46.2~\mathrm{k}\Omega \pm 41.2~\Omega$, whereas the 200~$\mu$m~s$^{-1}$ wire shows $2.68~\mathrm{M}\Omega \pm 836~\Omega$, corresponding to an $\sim 58\times$ difference (see Figure~\ref{fig:4}b for complete set of conductivity measurements). Details of the graphitic pads are provided in Supporting Note 1. Under optical illumination, the laser-written track appears dark relative to the surrounding single-crystal diamond. This contrast arises because conversion from transparent, wide-band-gap sp3 diamond to disordered sp2 carbon markedly increases visible-wavelength absorption and slightly raises the refractive index, thereby reducing transmittance through the modified region~\cite{prelas1997handbook,Wang2000}. Therefore, the degree of darkening can be used as a first qualitative indication of graphitization. However, it conflates absorption, scattering, debris, and illumination artifacts, and it is not phase-specific. For this reason, PL and hyperspectral Raman mapping are regarded as more reliable approaches, since they provide phase-specific, quantitative, and spatially resolved indicators of graphitization.

In the PL maps (Figure~\ref{fig:1}b,c), the laser-written graphitized regions appear dark because the processing introduces non-radiative recombination pathways that quench the broadband fluorescence. Yet, it is not phase-specific and can be easily affected by broadband emission, as shown in Figure~\ref{fig:1}c and detailed in Supporting Note 2. 
On the other hand, Raman mapping provides phase-specific contrast. 
Local graphitization converts sp3 diamond to sp2 carbon, depleting the sp3 line and enhancing the G band; accordingly, we construct sp3 maps by integrating 1325–1340~cm$^{-1}$ and sp2 maps by integrating 1575–1610~cm$^{-1}$~\cite{prelas1997handbook,Kononenko2008}. These phase-specific features (sp3 line, sp2 band, D band) enable quantitative, cross-structure comparison of phase content, crystallinity, and defect populations in laser-written carbon~\cite{Sotillo2018,Kononenko2008,Ashikkalieva2019}. Note that, as shown in (Figure~\ref{fig:1}c), Raman signal can also be contaminated by non-specific fluorescent background, an effect that can oftentimes be minimized with baseline correction and other techniques \cite{wei2015review}.


\subsection{Spectral characterization of Laser-written Electrodes}
\label{sec:Spectroscopic Study}

As shown in Figure~\ref{fig:1}, one of the limitations of both brightfield microscopy and broadband PL is that despite being good techniques for qualitative assessment of the fabrication they present limitations when used to provide phase-specific quantitative information. In the case of PL, it can in principle be used for quantitative sp3-to-sp2 conversion evaluation, but its sensitivity to broadband emission not stemming from the process of interest reduces its fidelity. This is further discussed in Supporting Notes 3 and 4. Therefore, we assessed the suitability of Raman spectroscopy for this task.

In graphitic carbons, the in-plane first-order $E_{2g}$ phonon (“G band”) appears near $\sim1580~\mathrm{cm^{-1}}$, while disorder or finite crystallite size activates the defect-induced “D band” around $\sim1350{-}1360~\mathrm{cm^{-1}}$\cite{Tuinstra1970,Ferrari2000}. In single-crystal diamond, the zone-center optical phonon of $F_{2g}$ symmetry gives a sharp first-order line at $1332\mathrm{cm^{-1}}$\cite{Knight1989,Ferrari2004}. 
These emphasize the phase origin of each peak and facilitates quantitative comparison across laser-written structures, enabling identification and quantification of carbon phases, crystallinity, and defect populations~\cite{Sotillo2018,Kononenko2008,Ashikkalieva2019}.

\subsubsection{Single-band characterization}
\begin{figure}[H]
    \centering
    \includegraphics[width=\linewidth]{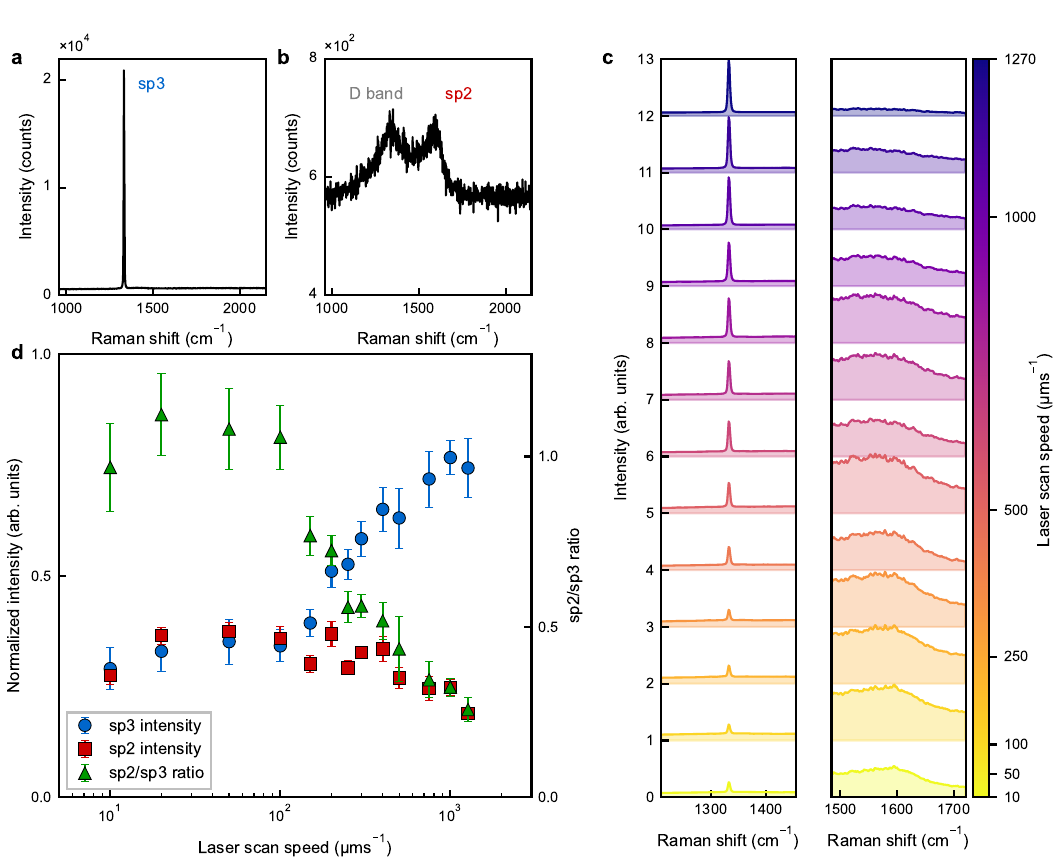}
    \caption{\textbf{Spectroscopic analysis of laser-written graphitic structures at different scan speeds.}~\textbf{a-b)} Representative normalized Raman spectra (532~nm excitation) from pristine diamond and a fully graphitized region (see \nameref{sec:Method} and Supporting Note 2 for normalization and pixel selection). The pristine diamond spectrum is dominated by the sp3 line at 1332~cm\(^{-1}\). In the graphitized region the sp3 signal is strongly suppressed, while a broad disorder-activated D band appears near 1350~cm\(^{-1}\) together with the sp2 G peak centered around 1580~cm\(^{-1}\), indicative of local graphitization.~\textbf{c)} Scan-speed-dependent mean Raman spectra restricted to the sp3 (diamond) and sp2 (graphitic) windows, shown for wires written at 10–1270~$\mu$m~s$^{-1}$ (top to bottom). For each speed, the mean is computed from 60 spectra per wire extracted from the hyperspectral maps, revealing progressive depletion of sp3 features and growth of the sp2 response with decreasing laser scan speed.~\textbf{d)} Speed-dependent quantitative summary: integrated intensities (from 60 spectra per wire extracted from the hyperspectral maps) of the sp3 and sp2 windows, together with the sp2/sp3 ratio, plotted against scan speed, error bar indicating one sigma error.}
\label{fig:2}
\end{figure}

Using the integrated confocal microscope, we collected hyperspectral Raman maps on every graphitic wire and pad fabricated at distinct scan speeds (see \nameref{sec:Method}). 
Figures~\ref{fig:2}a and b display representative spectra for (a) the unmodified single-crystal diamond and (b) a fully graphitized pad fabricated at a laser scan speed of 100~$\mu$m~s$^{-1}$. The pristine crystal is dominated by the sp3 line at 1332 cm\(^{-1}\). In contrast, the graphitic structure written with low laser scan speed exhibits depleted sp3 intensity and is governed by the disorder-induced D band (1350~cm\(^{-1}\)) together with the sp2 band (graphitic G band (1580 cm\(^{-1}\))), evidencing almost complete conversion to sp2 carbon.

Figure~\ref{fig:2}c compiles the speed-dependent averaged spectra of the laser-written graphitic wires. The spectral data normalized across all the samples to the sp3 peak (left panel) and the sp2 band (right panel), allowing the two regimes to be compared on equal footing across writing speeds. At the highest velocity tested (500~$\mu$m~s$^{-1}$) the spectrum is sp3-dominated and the D and G bands are barely discernible, implying that only a small fraction of lattice sites have been converted to sp2. Progressively reducing the speed, and thereby increasing the deposited energy density per unit length, reverses this balance: the D and sp2 bands intensify while the sp3 line diminishes, signaling enhanced graphitization.
The evolution is summarized quantitatively in Figure ~\ref{fig:2}d, which plots the normalized integrated intensities \(I_{\mathrm{sp3}}\) and \(I_{\mathrm{sp2}}\) as well as their ratio \(I_{\mathrm{sp2}}/I_{\mathrm{sp3}}\) versus laser scan speed. \(I_{\mathrm{sp3}}\) depletes with decreasing speed, whereas \(I_{\mathrm{sp2}}\) increases, except for a modest downturn below 50~$\mu$m~s$^{-1}$, where excessive dose does not yield higher sp2 band intensity. 


\subsubsection{Multi-band characterization}
\label{sec:multiband}
Laser-written electrodes in diamond contain both unconverted sp3-bonded lattice and newly formed sp2-bonded carbon. Within a confocal voxel, these phases typically coexist as sub-micron domains, so a single Raman spectrum is an average over a heterogeneous local volume. 

In the previous section we showed that the degree of graphitization can be inferred from the intensity of certain bands (e.g., sp3 line). However, relying on a single peak can be fragile in practice because the apparent intensity can be influenced by various factors such as laser power fluctuations, defocusing, local topography and debris, baseline fluorescence, or strain-induced shifts/broadening. 

To mitigate these effects, ratio metrics that combine several informative bands are usually preferred over single-peak readouts. In our case the most natural choice is to use the sp2/sp3 ratio, containing the information carried by the G/D complex and the 1332~cm$^{-1}$ diamond line. Figure~\ref{fig:3} illustrates this for a representative wire written at 50~$\mu$m s$^{-1}$.
\subsubsection{Hyperspectral unmixing}
\label{sec:Spectral Unmixing}
\begin{figure}[H]
    \centering
    \includegraphics[width=\linewidth]{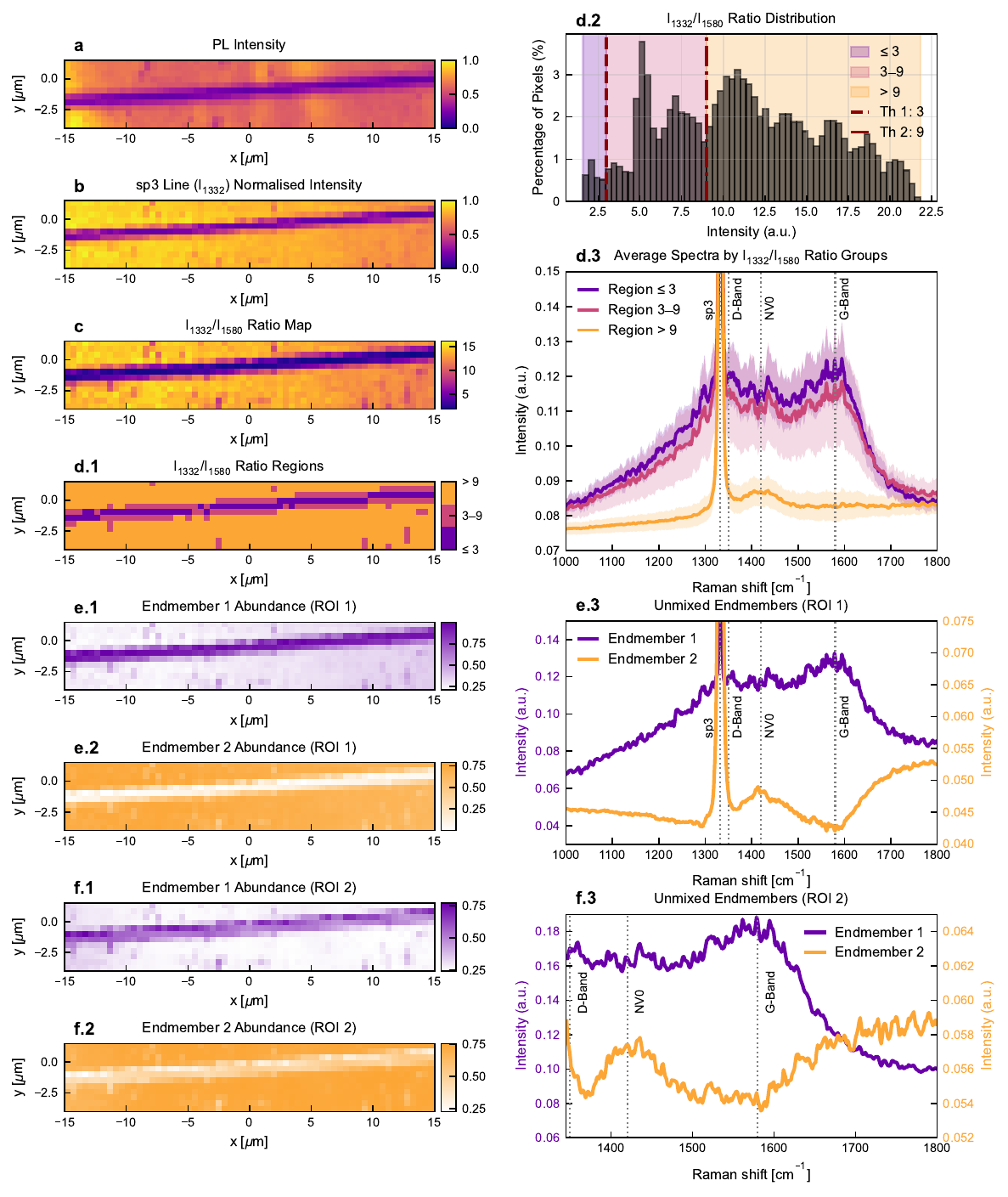}
    \caption{\textbf{Evaluation of spectral unmixing for electrode surface graphitization.}
\textbf{a-b)} PL intensity map and normalized first-order Raman intensity (sp3 line at 1332~cm$^{-1}$) for a graphitic wire written at 50~$\mu$m\,s$^{-1}$ (74.8~k$\Omega$).
\textbf{c)} Map of sp3-to-sp2 ratio. \textbf{d)} Threshold-based segmentation by the sp3-to-sp2 ratio $\rho = I_{1332}/I_{1580}$.
(\textbf{d.1}) Segmentation map using experimentally determined thresholds for non-graphitized ($\rho\leq3$), partially graphitized ($3<\rho<9$), and fully-graphitized ($\rho\geq9$).
(\textbf{d.2}) Pixel distribution of $\rho_{sp3}$ over the complete set of graphitic wires spectral data (i.e., including all different writing speeds), depicting selected thresholds.
(\textbf{d.3}) Class-averaged spectra (mean~$\pm$~1~s.d.) for the 74.8~k$\Omega$ electrode, showing the sp3 line (1332~cm$^{-1}$), D band ($\sim$1350~cm$^{-1}$), NV$^{0}$ fluorescence ($\sim$1420~cm$^{-1}$), and G band ($\sim$1580~cm$^{-1}$).
\textbf{e)} Spectral unmixing by Vector Component Analysis (VCA) with Fully Constrained Least Squares (FCLS) over spectral region of interest (ROI) 1 (1000-1800cm-1): FCLS abundance maps for endmember~1 (\textbf{e.1}) and endmember~2 (\textbf{e.2}), and corresponding endmember spectra with band markers (\textbf{e.3}).
\textbf{f)} Same unmixing workflow over ROI~2 (1345–1800~cm$^{-1}$, excluding the 1332~cm$^{-1}$ line): abundance maps for endmember~1 (\textbf{f.1}) and endmember~2 (\textbf{f.2}), and endmember spectra (\textbf{f.3}).
}
\label{fig:3}
\end{figure}

While ratio maps remove much of the acquisition variability, they still require manual selection of specific bands. Therefore, we also introduce the use of hyperspectral unmixing as a label-free alternative for fabrication processes where the bands of each phase might not be known. 
As described in \nameref{sec:Method}, we model each spectrum as a linear mixture of $n$ endmembers  and use well-established linear unmixing methods, such as vector component analysis (VCA)~\cite{nascimento2005vertex} to estimate these endmembers directly from the data. Then, abundances are estimated per pixel solving the linear inverse problem. In particular, for the results shown in Figure \ref{fig:3}, we impose $n=2$ endmembers and enforce both non-negativity, and the sum-to-one of the abundances at every pixel using fully constrained least squares (FCLS)~\cite{heinz2001fully}, which enables a direct fractional interpretation without the need for external references.

As shown in Figure \ref{fig:3}e, when applied over a broad region of interest (1000–1800cm$^{-1}$) unmixing yields endmembers that, while not identical to the actual reference spectra presented in Figure~\ref{fig:3}), preserve the expected relevant features: the diamond-associated endmember is characterised by a sharp 1332~cm$^{-1}$ line, whereas the graphite-associated endmember exhibits a broad D/G complex with a suppressed diamond line. Additionally, the diamond-associated endmember also features a broader band centered around $\sim1430$~cm$^{-1}$, which is consistent with the zero-phonon line of neutral single nitrogen-vacancy defects (NV$^{0}$) in diamond~\cite{Manson2013}. The resulting abundance maps also closely track the simpler sp3-window intensity maps (Figure \ref{fig:3}b).

To test sensitivity to the dominant diamond line, we repeated the analysis after excluding 1332~cm$^{-1}$ (spectral ROI = 1345–1800~cm$^{-1}$). In this case, the diamond-associated endmember is defined primarily by the NV$^{0}$ contribution within the spectral ROI, while the complementary endmember remains the broad G/D complex. The corresponding abundance maps retain the expected spatial structure but exhibit lower signal-to-noise (SNR), as anticipated when the most intense, phase-specific line is removed. 

Overall, the unmixing results reinforce the interpretation from ratio maps while removing the need to hand-pick bands: knowing only that two materials are present, the algorithm discovers representative endmembers and delivers per-pixel abundance maps that agree with the PL/Raman contrasts.

\subsection{Raman Monitoring of Electrical Performance}
\label{sec:Electrical Performance}
Ultimately, the optimal feedback signal for in-situ monitoring of fabrication should be directly tied to the functional performance of the resulting devices. For graphite electrodes on diamond, this performance is best represented by the conductivity of the laser-written features. To this end, the electrical conductivity of the pads and wires was measured using a custom setup (see \nameref{sec:Method}). 

\begin{figure}[H]
\centering
\includegraphics[width=\linewidth]{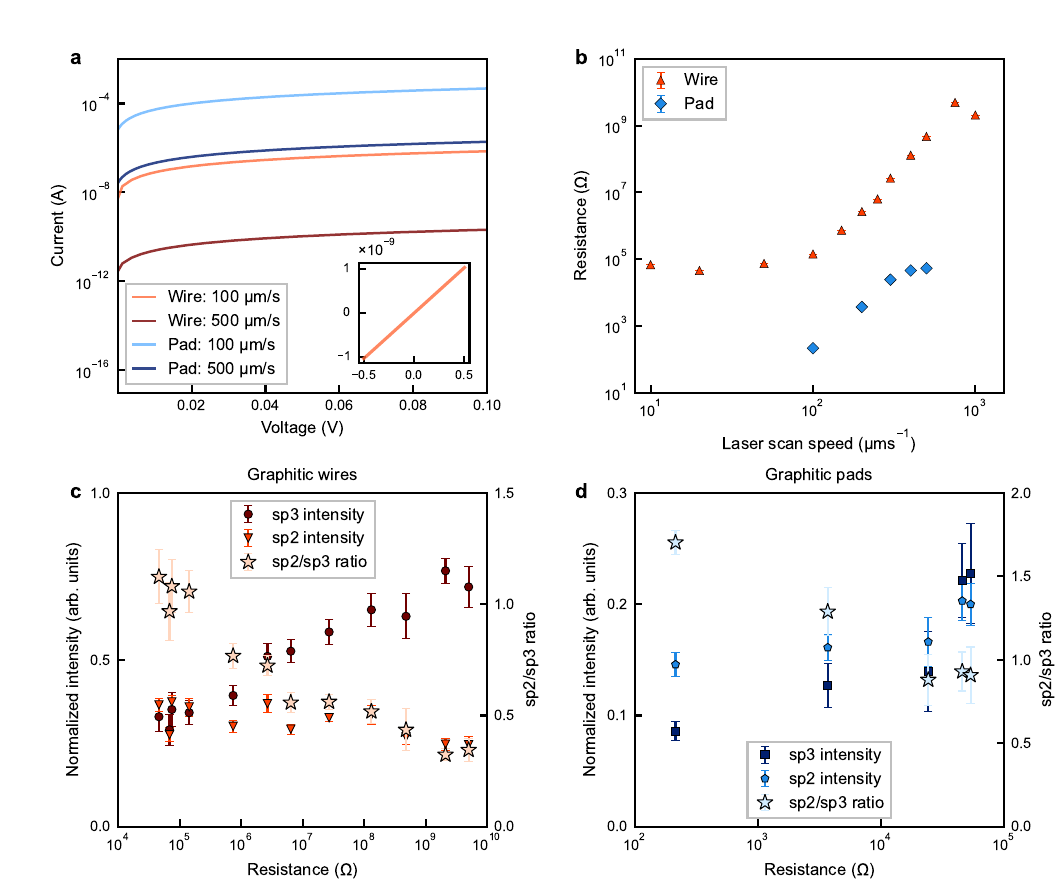}
\caption{\textbf{Direct Raman-electrical correlation of fs-laser-induced graphitization.}~\textbf{a)} Semi-log $I$--$V$ curves from 0 to 100~mV for pads and wires written at scan speeds of 100 and 500~$\mu$m~s$^{-1}$ (100~$\mu$m~s$^{-1}$: light red = wire, light blue = pad; 500~$\mu$m~s$^{-1}$: dark red = wire, dark blue = pad). Inset: linear-scale $I$-$V$ for a representative wire over $-500$ to $+500$~mV, showing symmetric ohmic behavior. Pad electrodes generally exhibit lower resistance than wires, attributable to larger cross-sections and a higher effective degree of graphitization.~\textbf{b)} Resistance of laser-written pads and wires versus laser scan speed. Slower scan speeds yield lower resistance. Values are obtained from linear least-squares fits to I–V data acquired within \(\pm 500\)~mV in a two-probe configuration (see \nameref{sec:electrical_conduct_meas}), error bar indicates one sigma error. Of note, at the lowest scanning speeds, wire resistance is comparable to that of the contact electrodes, reducing measurement accuracy. (see \ref{sec:Method}). Also, the wire written at $750~\mu$ms$^{-1}$ exhibits non-ohmic $I$–$V$ behavior (see Supporting Note~10), giving a higher apparent resistance than the wire written at $1000~\mu$ms$^{-1}$. ~\textbf{c-d)} Resistance dependence of the mean normalized Raman intensities in the sp3 (diamond) and sp2 (graphitic) windows, as well as the sp2 to sp3 ratio, is shown.}
\label{fig:4}
\end{figure}

Figure~\ref{fig:4} presents our findings and consolidates the link between laser scan speed, local bonding state, and device-level conductivity of the laser-written features. Figure~\ref{fig:4}a depicts the current--voltage traces for all wires and pads, which are linear over $\pm 500$~mV, indicating ohmic behavior and no measurable Schottky barriers; conduction is therefore governed by the volume fraction and connectivity of the laser-generated graphitic phase. For graphitic structure of width of 200~$\mu$m written at $100~\mu$ms$^{-1}$, the pad resistance is $332~\Omega \pm 0.1~\Omega$, versus $1.46\times10^{5}~\Omega \pm 48.8~\Omega$ for the wire at the same speed. The lower pad resistance reflects its larger cross-section and additional over-writing during fabrication, which further graphitizes the diamond and increases the graphite fraction.
All resistances are extracted from linear fits to quasi-static I–V sweeps within \(\pm 500\)~mV (details of the IV curves are shown in Supporting Note 10 and 11).

In Figure~\ref{fig:4}c and d we plot, separately for wires and pads, the mean normalized integrated sp3 and sp2 intensities as well as their ratio versus laser scan speed. 
As the scan speed increases, the resistance rises concurrently with the normalized sp3 intensity, indicating reduced graphitization and weakened conductive pathways. For wires (Fig.~\ref{fig:4}c), decreasing the scan speed from high to intermediate increases the normalized sp2 intensity and lowers the resistance. However, once the sp3 contribution is strongly diminished (deeply graphitized regime), sp2 intensity alone becomes a less reliable measure of further conductivity enhancement. Since the sp2 intensity rises as the sp3 intensity decreases with scan speed, their ratio mirrors these trends. Pads show the same qualitative behavior (Figure~\ref{fig:4}d) but remain systematically more conductive than wires at identical scan speeds.

\subsection{Discussion}


Figure~\ref{fig:2} summarizes how the Raman features evolve with writing dose, controlled by scan speed. Reducing the speed, which increases the energy deposited per unit length, produces a monotonic depletion of the 1332~cm$^{-1}$ sp3 line and a general rise of the $\sim$1580~cm$^{-1}$ sp2 band, consistent with progressive diamond-to-graphite conversion. At the slowest speeds, however, the sp2 intensity reaches a maximum and then declines despite continued sp3 depletion, a behavior also reported in other studies on diamond graphitization~\cite{Ali2022,Khomich2023}.

This counter-intuitive reduction in sp2 intensity can be interpreted within the ‘amorphization trajectory’ framework of \textit{Ferrari et al.}~\cite{Ferrari2000,Ferrari2004}, wherein progressive disorder drives a transition from nanocrystalline graphite to amorphous sp2-rich carbon and, ultimately, to highly disordered or tetrahedral amorphous forms. In this regime, shrinking and fragmenting sp2 clusters suppress well-defined Raman modes, so the apparent sp2 signal diminishes even as the actual sp2 bonding fraction remains high.
At intermediate doses, ordered sp2 clusters grow, strengthening the D and G bands. At extreme doses, however, high defect densities and reduced cluster sizes suppress well-defined vibrational modes, broadening and merging the D and G peaks into a low-contrast background. The resulting spectral degradation reduces the apparent sp2 Raman intensity, even when the actual sp2 bonding fraction remains high. Therefore, under such high-dose conditions, sp3 depletion provides a more reliable metric for tracking graphitization progression, as it decreases monotonically with increasing laser dose, is insensitive to spectral contrast loss, and, thanks to its higher SNR, is less sensitive to non-specific broadband background \cite{Ali2022,Wang2000}.

Pad writing shows stronger sp3 depletion than single-pass wires, which we attribute to stress-assisted graphitization during overwriting. Lattice mismatch generates tensile stress in the transformed layer and in-plane compressive stress in adjacent diamond~\cite{Wang2000}. Compression lowers the graphitization barrier, so overlapping passes enhance conversion. The observed wavenumber shift of the diamond line is consistent with established stress coefficients~\cite{Grimsditch1975,Zaitsev2001,Orwa2000} (see Supporting Note~6) and explains the lower resistance measured for pads at identical scan speeds.

In Figure \ref{fig:3} we introduce the use of more advanced Raman-based structure-characterization metrics. Particularly we first present multi-band metrics, mainly sp3-to-sp2 ratio. Using this ratio has the advantage of being agnostic to some of the effects that can generate non-structure-specific intensity fluctuations in the sp2 and sp3 peaks, such as excitation power. Moreover, we also introduce hyperspectral unmixing as a label-free complementary spectral analysis (Fig.~\ref{fig:3}). A linear mixing model with non-negativity and sum-to-one constraints recovers two data-driven endmembers that map onto diamond-like and graphite-like responses, producing phase-fraction maps that agree with band-based indicators while leveraging the full spectrum. The approach is advantageous as it does not rely on pre-established bands of interest and is also robust to power drift and can potentially help reduce the effect of baseline fluorescence, though the linearity assumption presents some limitations (see \nameref{sec:Method}). Yet, in the case of monitoring graphitization in diamond, where the bands of interest are well-known, hyperspectral unmixing does not provide a significant advantage.

Finally, in Figure~\ref{fig:4} we show how Raman spectroscopy can be used to evaluate the performance of graphite electrodes. 
Across devices, depletion of the 1332~cm$^{-1}$ sp3 line correlates with resistance more robustly than absolute sp2 intensity. This is likely due to a combination of the sensitivity of sp2 to disorder in high-dose conditions and the effect of the broadband background from nongraphitic damage, which is more severe in sp2 in the low-dose regime due to the difference in SNR with the sp3 line (Fig.~\ref{fig:1}, Supporting Note~2). 
As shown in Figures \ref{fig:4}c,d, the sp2-to-sp3 ratio is a useful indicator of graphitization because relative normalization reduces sensitivity to power fluctuations. Yet, under high-dose irradiation, sp3 exhibits less variance than sp2-to-sp3 ratio, likely due to the effect of extreme disorder on the sp2 band~\cite{Ali2022,Wang2000}.

\section{Conclusion and Outlook}

Raman microspectroscopy provides a phase-specific, quantitative readout that links to ultrafast laser writing conditions and device performance in diamond. Across surface pads and wire electrodes, depletion of the 1332~cm$^{-1}$ sp3 line predicts conductivity more reliably than absolute sp2 and PL intensities. Some of these deficiencies could be solved via spectral pre-processing, which has the drawback of adding complexity, time, and additional sources of error \cite{chen2018fast}.

Therefore, a self-referenced sp3 metric, benchmarked to the local pristine background, is therefore a robust in situ indicator, while the sp2/sp3 ratio remains useful at moderate doses since it is less sensitive to changes in experimental conditions. Thus, Raman microspectroscopy can be integrated into the fabrication system to provide real-time, in-situ feedback of the device fabrication. In this integrated application, rather than generating the maps shown in this work, the excitation laser could be spatially offset with a lag with the fabrication laser and sample scattering could be collected with sub-second integration times. Additionally, as shown in Supporting Note~2, the excitation beam would contribute to removing surface debris.

As part of this work, linear hyperspectral unmixing has been introduced as a full-spectrum, label-free alternative for post-fabrication mapping and indicator validation. In the case of diamond surface graphitization, unmixing performs analogously with band-based metrics, demonstrating its potential for samples with less defined structure-specific bands. 

The methodology presented in this work is not specific to diamond structure indicator and could therefore be adapted to other hosts and functionalities (e.g., fabrication in glasses and wide-bandgap crystals~\cite{fernandez2014role,Li2008,Little2010}, laser-induced graphene on polymers~\cite{Wang2025}, and embedded heaters or electrodes in SiC, sapphire, silica, and ceramics~\cite{Choi2016,Naseri2022,Zhang2025}), providing a practical route toward specification‐driven fs‐laser microfabrication.

\section{\label{sec:Method}Materials and Methods}
\subsection{Electrode fabrication}

A single-crystal, optical-grade $\langle100\rangle$ diamond grown by chemical-vapor deposition (Element Six; nominal N concentration $<\!1$~ppm) served as the substrate for this study.  
Surface graphitic pads and wires were inscribed with a fs laser operating at 520~nm ($\tau_\text{p}=400$~fs, $f_\text{rep}=10$~kHz,) and a pulse energy of 30~nJ, using various laser scan speed. Wave-front aberrations were corrected by a liquid-crystal spatial light modulator, and the beam was focused through an objective lens with numerical aperture $\mathrm{NA}=0.50$ (ZEISS Plan-Neofluar, 0.5NA, 20x).  
Wire electrodes were written in a single pass under continuous stage translation (no overwriting). Successive pulses partially overlapped to form a continuous track; at fixed $f_\text{rep}$ the pulse-to-pulse separation was set by the scan speed. Pad electrodes were produced by stacking parallel wires with a constant 0.5~$\mu$m pitch, maintained for all scan speeds.

\subsection{\label{sec:electrical_conduct_meas}Electrical conductivity measurement}
The resistance of laser-written pads and wires was determined from two-probe current–voltage (I–V) measurements using a low-noise transimpedance amplifier (DLPCA-200, FEMTO) together with a real-time voltage source and data-acquisition controller (ADwin-Gold II, Jäger). Electrical contact was made with two micromanipulated Pt probes. For pad devices, the probes contacted opposite ends of each laser-written pad. For wire devices, both termini were first connected to laser-written square pads (50~$\mu$m~\(\times\)~50~$\mu$m, written at 100~$\mu$m\,s\(^{-1}\)), and the probes landed on these pads to standardize the contact area. The ADwin generated quasi-static voltage sweeps within the ohmic window (typically \(\pm 500\)~mV) and synchronously digitized the DLPCA-200 output.

\subsection{Brightfield imaging}
Optical micrographs were captured with an integrated transmission optical microscope. A red-LED array situated under the sample serves as illumination source and the images are captured on a CCD camera (EC650, Prosilica Inc.). 

\subsection{Confocal Imaging}
Our fabrication setup has an integrated custom confocal microscope that can be simultaneously utilized for photoluminescence and Raman spectral collection. Using a 532~nm continuous-wave excitation laser (Cobolt Samba 150mW) and a fast steering mirror (FSM-300, Newport) relayed via a 4f into the back aperture of the system's objective, 2D photoluminescence and Raman maps can be collected. Notably, in this study, the 2D maps were collected post-fabrication utilizing a higher numerical aperture objective (Olympus PlanApo 0.95NA, 60x). 

Both photoluminescence and Raman spectra were collected simultaneously employing 30~mW of laser excitation. The samples were scanned with a 500nm pixel-to-pixel lateral displacement and a 1 second exposure time. The collected signal was split 99:1 between the Raman and the photoluminescence collection paths. For the latter, signal was collected via a single photon avalanche detector (SPCM-AQRH-14-FC, Excelitas Inc.). Given that the collection path of our confocal setup features a 550~nm long-pass dichroic, and that the SPAD quantum efficiency drops significantly after 850~nm, the PL collection window can be considered 550-850~nm. For the Raman collection path, scattered light was dispersed by a 1200~lines~mm$^{-1}$ blazed spectrograph (SpectraPro HRS-1200) and detected on a deep-cooled CCD camera (PIXIS 100, Princeton Instruments). 

\subsection{Raman spectra preprocessing}

The hyperspectral Raman data were loaded as a three-dimensional array $I(x,y,\lambda)$. To enable cross-dataset comparison (accounting for sample tilt, focus drift, and excitation-power fluctuations), we pre-normalized each map using pristine diamond regions identified from the sp3 window. Specifically, we integrated $I(x,y,\lambda)$ over $[1325,1340]$~cm$^{-1}$ to obtain $I_{\mathrm{int}}(i,j)$, ranked all pixels, and defined the $N_{\mathrm{bright}}$ brightest pixels (wire electrodes: $N_{\mathrm{bright}}=200$; pad electrodes: $N_{\mathrm{bright}}=1000$) as the pristine reference. The normalization factor $F_{\mathrm{norm}}$ was the mean of these $I_{\mathrm{int}}$ values, and the hyperspectral cube was normalized as $I_{\mathrm{norm}}(i,j,\lambda)=I(i,j,\lambda)/F_{\mathrm{norm}}$, establishing $\langle I^{\mathrm{norm}}{\mathrm{sp3}}\rangle{\mathrm{pristine}}=1$.

From $I_{\mathrm{norm}}$, we computed integrated intensities for sp3 (diamond, $[1325,1340]$~cm$^{-1}$) and sp2 (graphitic, $[1575,1610]$~cm$^{-1}$) at each pixel and identified $N_{\mathrm{dark}}$ darkest sp3 pixels ($60$ for wire and $1500$ for pad) as modified regions. For these pixels, we report the spatial coordinates, the normalized integrals $I^{\mathrm{norm}}{\mathrm{sp3}}$ and $I^{\mathrm{norm}}{\mathrm{sp2}}$, their averages, and the ratio $R=I^{\mathrm{norm}}{\mathrm{sp2}}/I^{\mathrm{norm}}{\mathrm{sp3}}$ as a quantitative metric of graphitization. Fixed integration windows were used throughout (sp3 width $15$~cm$^{-1}$; sp2 width $35$~cm$^{-1}$); thus, the sp2/sp3 ratio is the ratio of integrated band intensities, not intensity per unit wavenumber. Full algorithmic details, parameter choices, and examples are provided in the SI (see Supporting Note 3).
\subsection{Spectral Unmixing}
\label{sec:unmixing}

In this work we adopt the widely used \emph{linear mixing model} (LMM), in which each spectrum $x \in \mathbb{R}^{b}$ is expressed as a linear combination of $n$ endmember spectra $m_{i} \in \mathbb{R}^{b}$:
\begin{equation}
    x = \sum_{i=1}^{n} \alpha_{i} m_{i} + \varepsilon 
\end{equation}
where $\alpha_{i}$ denotes the abundance of the $i$-th endmember, and . 

\paragraph{Endmember identification.} 
To identify the endmember spectra from the data, we employed two well-established blind unmixing algorithms (a)\textit{Vertex Component Analysis (VCA)}~\cite{nascimento2005vertex}, a fast geometrical method that projects the dataset along random directions to iteratively identify extreme points of the spectral simplex, which serve as candidate endmembers; and (2) \textit{N-FINDR}~\cite{winter1999n}, an algorithm that searches for the set of endmembers maximising the volume of the simplex formed by candidate spectra, under the assumption that pure pixels are present in the dataset. Both algorithms require specifying the number of endmembers in advance. Here, we constrained the problem to two endmembers (sp2 and sp3), which reflects the known physics of the system and avoids overfitting.

\paragraph{Abundance estimation.}
Once endmembers are identified, abundances can computed by solving the linear inverse problem under different constraints: (a) \textit{Fully Constrained Least Squares (FCLS)}~\cite{heinz2001fully}: enforces both non-negativity ($\alpha_{i} \geq 0$), and the sum-to-one constraint ($\sum_{i=1}^{n} \alpha_{i} = 1$), ensuring abundances can be interpreted as fractions; and (b) \textit{Non-Negative Least Squares (NNLS)}~\cite{lawson1995solving}: relaxes the sum-to-one condition but maintains non-negativity, which can be useful when baseline variations are present. As shown in Figure \ref{fig:3}, the abundance maps at 50~$\mu$m/s, the pristine diamond region does not show an abundance of exactly 0 for the graphite endmember and 1 for the diamond endmember. This is expected since the endmembers are extracted across all wire maps, focus and intensity differences between datasets also shift the estimated simplex (see Supporting Note 8). The result is a small residual assignment of sp2 abundance even in pristine regions. Importantly, the maps still provide clear contrast: pristine areas are dominated by the sp3 endmember, while graphitized tracks are dominated by sp2, confirming the physical validity of the decomposition.

\paragraph{Limits of linear mixing model.}
Raman scattering is, to first order, linear in the number of scatterers, so spatial coexistence of sp3 and sp2 domains within the confocal volume leads to additive spectra, which the LMM captures. Moreover, our maps are recorded on polished surfaces with modest absorption at 532~nm outside the most heavily graphitized regions, reducing multiple scattering and re-absorption that could induce non-linear mixing. That said, three effects can challenge strict linearity: (i) strong broadband fluorescence in highly damaged zones, (ii) self-absorption and re-emission in thick, strongly absorbing tracks, and (iii) spectral distortions from stress/strain gradients that shift or broaden peaks in a way not representable by a fixed endmember. In such cases, \emph{non-linear} or \emph{bilinear} models (e.g., polynomial post–nonlinear mixing, kernel-based unmixing, or autoencoder-based unmixing \cite{georgiev2024hyperspectral}) may yield incremental gains, at the cost of additional parameters and reduced interpretability. 

\paragraph{Implementation.} We conducted the hyperspectral unmixing analysis with Python using the RamanSPy package \cite{georgiev2024ramanspy}.

\section{Acknowledgments}  This work was supported by the UK Engineering and Physical Sciences Research Council (EP/W025256/1).

\section{Disclosures}
The authors declare no conflicts of interest.

\section{Data availability} The data that support the findings of this study are available from the corresponding authors, [X.C., A.F.G.], upon reasonable request.


\newpage
\setcounter{figure}{0}        
\renewcommand{\thefigure}{\arabic{figure}}
\captionsetup[figure]{name=Supporting Figure}

\begin{center}
\Large\textbf{Supporting Information}
\end{center}

\section*{Supporting Note 1: Pad optical micrograph and Raman map}

\begin{figure}[H]
\centering
\includegraphics[width=0.9\textwidth]{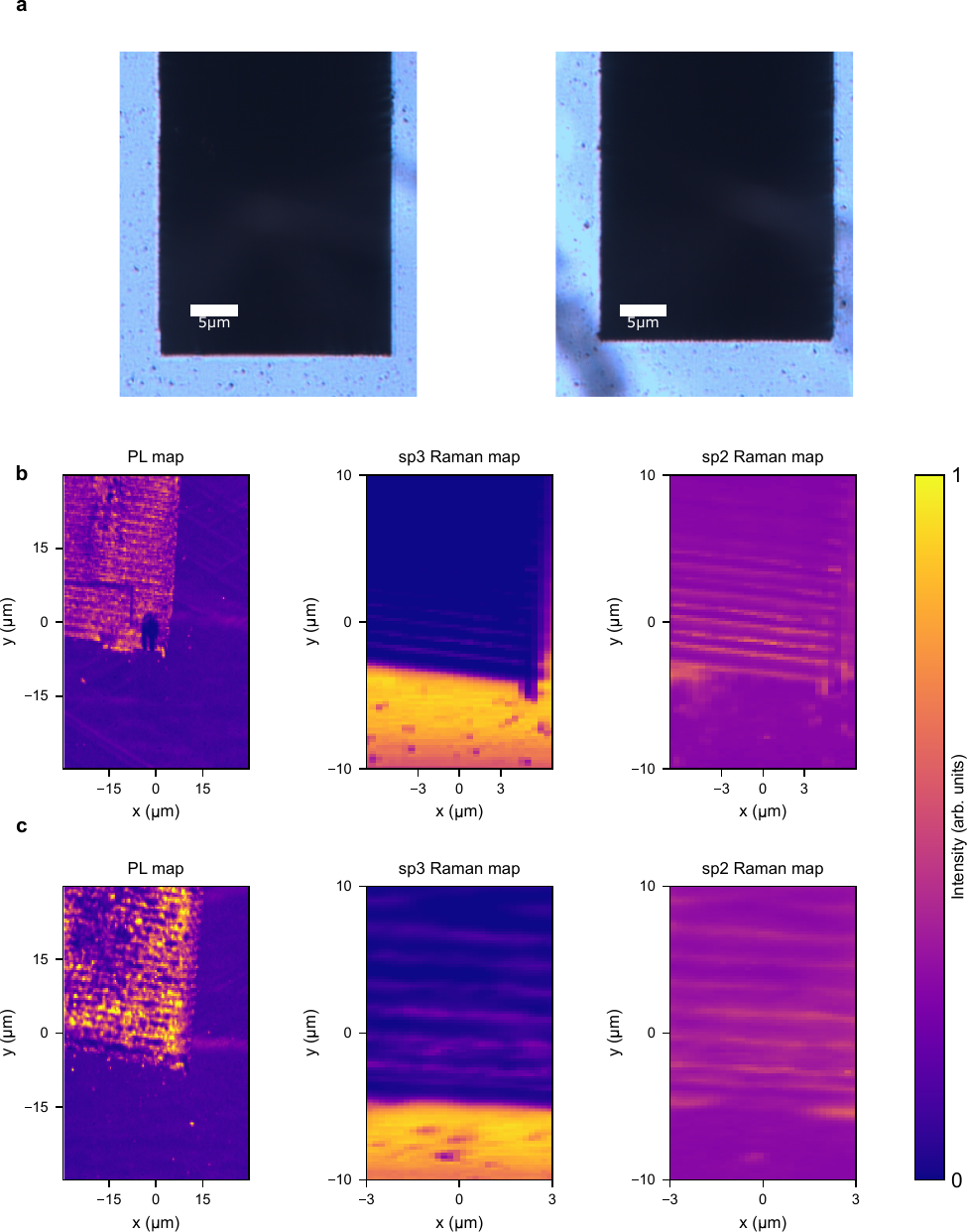}
\caption{\textbf{Laser-written graphitic structures with optical/PL/Raman characterisation.}\textbf{a-c)} Optical micrographs, PL maps, and Raman intensity maps (sp3 and sp2 windows) for \emph{pads} written at scan speeds of 100 and 200~$\mu$m~s$^{-1}$, respectively. The 100~$\mu$m~s$^{-1}$ pad exhibits a resistance of $214~\Omega \pm 10.4~\Omega$, whereas the 200~$\mu$m~s$^{-1}$ pad exhibits $3.69~\mathrm{k}\Omega \pm 1.59~\Omega$, i.e., a difference exceeding one order of magnitude. The optical darkening and PL contrast are effectively indistinguishable between the two speeds; thus, optical micrographs and PL maps cannot resolve electrodes that differ in resistance by more than an order of magnitude. By contrast, Raman mapping of the sp3 window provides a direct and more reliable measure of the degree of graphitization, offering phase-specific insight that complements the broadband PL response and the lower-contrast sp2 map.}\label{fig:S1}
\end{figure}

As described in the manuscript, we employed laser-written graphitic pads and wires to emulate two common classes of conductive elements. Supporting Figure~\ref{fig:S1}a shows an optical micrograph of pad electrodes with a width of 50~$\mu$m and length of 200~$\mu$m, written at scan speeds of 100 and 200~$\mu$m~s$^{-1}$, respectively. The 100~$\mu$m~s$^{-1}$ pad exhibits a resistance of $214~\Omega \pm 10.4~\Omega$, whereas the 200~$\mu$m~s$^{-1}$ pad shows $3.69~\mathrm{k}\Omega \pm 1.59~\Omega$, i.e., a difference of more than an order of magnitude. Under optical illumination, the laser-written tracks appear dark relative to the surrounding single-crystal diamond. This contrast arises because converting transparent, wide-band-gap sp3 diamond to disordered sp2 carbon increases visible-wavelength absorption and slightly raises the refractive index, thereby reducing transmittance through the modified region~\cite{prelas1997handbook,Wang2000}. 
Supporting Figures~\ref{fig:S1}b and c present the corresponding photoluminescence (PL) and Raman (sp3 and sp2) maps. In both pads, the written regions appear bright in PL immediately after fabrication. This behavior differs from the wires, which appear dark in PL due to laser-induced non-radiative recombination pathways. 

The bright pad emission consists of locally activated NV$^0$ and NV$^-$ centers in this Type~Ib diamond (high nitrogen content), and other defects centers that producing a broad band that photo-bleaches under green (532~nm) excitation. This highlights that monitoring the full-window PL signal is not an ideal metric during fabrication, as relevant phase information is obscured by unrelated emissions and does not directly reflect device performance. In contrast, hyperspectral Raman mapping provides more specific structural information. By integrating the sp3 diamond window (1325-1340cm$^{-1}$)~\cite{Kononenko2008} and the sp2 G-band window (1575-1610cm$^{-1}$)~\cite{prelas1997handbook}, the sp3 map shows the laser-written regions as decreased peak intensity relative to pristine diamond, with stronger depletion for the lower scan speed. The sp2 map exhibits the expected behavior (i.e., increased peak intensity in the written region) but has lower overall contrast. Both the sp3 and sp2 Raman maps exhibit horizontal line-like features, which arise from the writing process. Such features are more pronounced in the sp2 than in the sp3 maps. Details explanation will be discussed in Supporting Note 2.

\begin{figure}[H]
\centering
\includegraphics[width=0.6\textwidth]{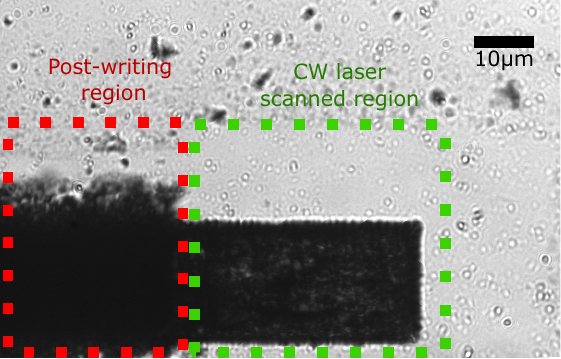}
\caption{\textbf{CW-laser-assisted removal of graphitic debris from laser-written pads.}
Optical micrograph of a graphitic pad after femtosecond (fs) laser writing. The dashed red box highlights debris generated during fs writing; the dashed green box marks a region subsequently scanned with a CW laser, where the surface graphitic debris has been effectively removed, revealing the underlying written structure.}\label{fig:S1_b}
\end{figure}

Supporting Figure~\ref{fig:S1_b} shows an optical micrograph of a pad acquired after PL mapping. Scanning a $>$20~mW, 532~nm beam around the electrode removes the initial femtosecond-laser-induced graphite debris~\cite{Wang2000}, revealing a clean structure (dashed green square) compared to the post-writing state (dashed red square). Thus, maintaining the 532~nm probe during Raman acquisition could serve the dual role of (a) enabling spectroscopy monitoring, and (b) clearing graphite debris. 

\section*{Supporting Note 2: Origin of contrast in hyperspectral maps}
\label{sec:SI2}

\begin{figure}[H]
\centering
\includegraphics[width=1\textwidth]{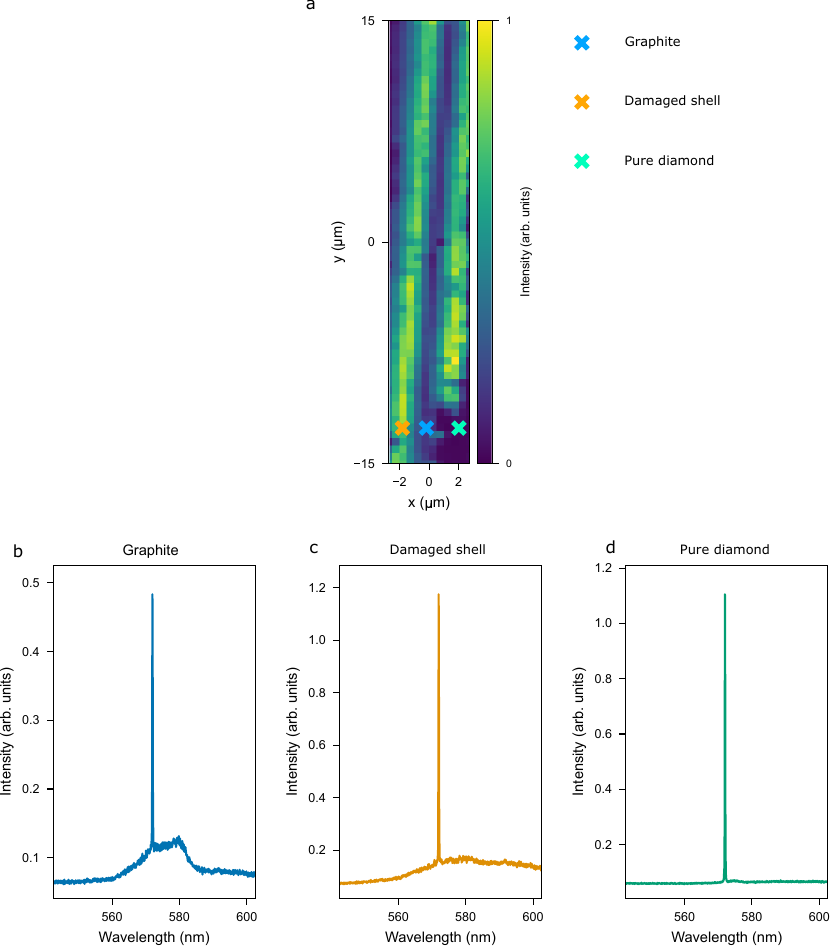}
\caption{\textbf{Raman hyperspectral analysis of a laser-written wire with surrounding damaged shell.}
~\textbf{a)} Raman map in the 590–600~nm range, excluding sp3 and sp2 bands. Bright regions surrounding the wire correspond to broadband background emission. Colored crosses mark representative pixels: blue for the graphite core, orange for the damaged shell, and light green for the pure diamond background.
~\textbf{b–d)} Self-normalized spectra from the selected pixels. The graphite spectrum (blue) shows depleted sp3, D, and sp2 features; the diamond spectrum (light green) is dominated by the sp3 line; while the damaged shell (orange) lacks D and sp2 bands but exhibits broad background emission attributed to defect activation.}\label{fig:S1_c}
\end{figure}

As shown in Supporting Figure~\ref{fig:S1_c}, both the sp3 and sp2 Raman maps exhibit horizontal line-like features, which are more pronounced in the sp2 map than in the sp3 map. The origin of these contrasts differs for the two cases. Since the pad is fabricated by stacking parallel wires with a 0.5~$\mu$m pitch, the bright features in the sp3 maps correspond to regions between adjacent wires where less damage occurs, resulting in reduced graphitization. In contrast, the bright features in the sp2 maps arise from a different mechanism, consistent with the behavior described in Figure~1c of the main text. Here, the increased brightness is not related to enhanced graphitization but instead originates from broadband background emission.

To confirm this, Supporting Figure~\ref{fig:S1_c} presents a Raman map in the 590–600~nm spectral range (the same hyperspectral map of Figure~1c in the main text but different spectral window), excluding the sp3 and sp2 bands. Compared to the laser-written wire and the pure diamond background, the surrounding damaged shell appears bright, confirming that the observed contrast does not originate from either the sp3 or sp2 Raman signals. Spectra from three representative pixels—assigned to the graphite core, the damaged shell, and the pure diamond background—are extracted in Supporting Figure~\ref{fig:S1_c}a and shown in Supporting Figure~\ref{fig:S1_c}b–d. Each spectrum is normalized for comparison. The graphite region exhibits depleted sp3, a clear D band, and sp2 features, while the pure diamond spectrum is dominated by the sp3 line. In the damaged shell, no D or sp2 bands are observed apart from the sp3 peak. Instead, a broad emission is present, which contributes to the increased brightness observed in both the photoluminescence and hyperspectral maps. This emission is attributed to defect activation in the damaged shell surrounding the wire core~\cite{Kononenko2008}.

\section*{Supporting Note 3: Pixel identification, integration, and normalization \label{sec:pixel_norm}}
\begin{figure}
\centering
\includegraphics[width=1\textwidth]{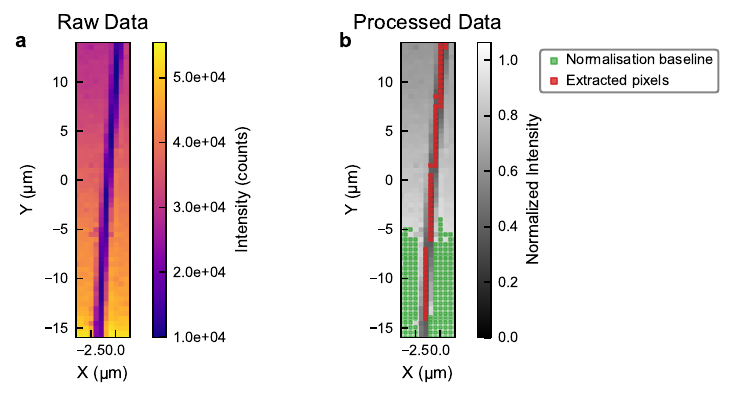}
\caption{\textbf{Normalization and pixel selection for hyperspectral Raman analysis.}~\textbf{a}, Raw Raman map integrated over the sp3 window $[1325,1340]$~cm\(^{-1}\).~\textbf{b}, Processed map after normalization to pristine diamond. Green pixels indicate the $N_{\mathrm{bright}}$ highest sp3-integrated intensities used to compute the normalization factor $F_{\mathrm{norm}}$ (baseline), and red pixels indicate the $N_{\mathrm{dark}}$ selected low-sp3 locations used for further analysis. The grayscale reports normalized sp3 intensity with the pristine-diamond baseline satisfying $\langle I_{sp3}^{norm}\rangle=1.0$. This procedure enables direct comparison of $I_{sp3}^{norm}$, $I_{sp2}^{norm}$, and the ratio $R=I_{sp2}^{norm}/I_{sp3}^{norm}$ across datasets. Axes are $X$-$Y$ in~$\mu$m.}
\label{fig:S2}
\end{figure}

The hyperspectral Raman data was loaded as a three-dimensional array $I(x, y, \lambda)$, where $x$ and $y$ represent the spatial coordinates, and $\lambda$ represents the wavelength. 
To compensate for experimental variations including focusing, and excitation laser power fluctuations between different datasets, we implemented a sample-specific normalization procedure based on pristine diamond regions (shown as green pixels in Fig.~\ref{fig:S2}). 

For the sp3 diamond peak analysis, we defined a wavelength window $\Lambda_{sp3} = [1325, 1340]$~cm\(^{-1}\). The integrated intensity for each pixel $(i,j)$ was calculated as:
\begin{equation}
    I_{int}(i,j) = \sum_{\lambda \in \Lambda_{sp3}} I(i,j,\lambda)
\end{equation}
For every sample, we identified the $N_{bright}$ brightest pixels (where $N_{bright} = 200$ for wire electrodes and $N_{bright} = 1000$ for pad electrodes) by sorting all pixels according to their integrated intensity values. These brightest pixels correspond to pristine diamond regions that serve as our normalization baseline (see Fig.~\ref{fig:S2} where the green pixels indicate the normalization baseline). The normalization factor was computed as the mean intensity of these brightest pixels:
\begin{equation}
    F_{norm} = \frac{1}{N_{bright}} \sum_{k=1}^{N_{bright}} I_{int}(p_k)
\end{equation}
where $p_k$ represents the $k$-th brightest pixel.

After establishing the normalization factor, the entire hyperspectral image was normalized:
\begin{equation}
    I_{norm}(i,j,\lambda) = \frac{I(i,j,\lambda)}{F_{norm}}
\end{equation}
Subsequently, we extracted and performed integrated intensity calculations for both the sp3 (diamond) and sp2 (graphitic carbon) spectral windows for those with dark sp3 intensity (red pixels in Fig.~\ref{fig:S2}). For each pixel, we computed:
\begin{equation}
    I_{sp3}^{norm}(i,j) = \sum_{\lambda \in [1325, 1340]} I_{norm}(i,j,\lambda)
\end{equation}
\begin{equation}
    I_{sp2}^{norm}(i,j) = \sum_{\lambda \in [1575, 1610]} I_{norm}(i,j,\lambda)
\end{equation}
From the normalized sp3 integrated intensity map, we identified the $N_{dark}$ darkest pixels (typically 60-2000 pixels depending on the sample types), which correspond to areas with reduced diamond features, indicating graphitization or structural modification.

The normalization procedure establishes a reference scale where a value of 1.0 represents the average integrated intensity of pristine diamond in the sp3 window (1325-1340~cm\(^{-1}\)). Mathematically, $\langle I_{sp3}^{norm} \rangle_{pristine} = 1.0$. This normalization enables direct comparison between different datasets and quantitative assessment of the depletion of sp3 Raman signal. For the darkest pixels identified in our analysis, $I_{sp3}^{norm} < 1.0$ indicates reduced diamond sp3 character relative to pristine regions, while $I_{sp2}^{norm}$ values can be directly compared to the sp3 baseline. The ratio $R = I_{sp2}^{norm}/I_{sp3}^{norm}$ provides a quantitative metric for the degree of graphitization, where $R \ll 1$ indicates predominantly diamond sp3 character and $R \geq 1$ suggests substantial sp2 carbon content.

For each identified dark pixel, we exported the normalized integrated intensities for both spectral windows along with their spatial coordinates. The average normalized intensity across all dark pixels was calculated as:
\begin{equation}
    \bar{I}_{window}^{dark} = \frac{1}{N_{dark}} \sum_{k=1}^{N_{dark}} I_{window}^{norm}(d_k)
\end{equation}
where $d_k$ represents the $k$-th darkest pixel and $window \in \{sp3, sp2\}$. This averaging procedure provides a statistical measure of the overall modification in the analyzed region, with the normalized scale ensuring that values can be directly compared across different measurements and sample positions. The exported data enables quantitative comparison of the degree of graphitization across different electrode configurations and processing conditions. We note that the sp2 integration window (15~cm\(^{-1}\)) is wider than the sp3 window (35~cm\(^{-1}\)), reflecting the intrinsic spectral widths of these features in diamond. As all measurements use identical integration windows, this systematic difference does not affect relative comparisons between samples. The sp2/sp3 ratio should therefore be interpreted as the ratio of integrated intensities over their respective characteristic spectral features, rather than intensity per unit wavelength.

\section*{Supporting Note 4: Quantitative PL analysis}

\begin{figure}[H]
\centering
\includegraphics[width=0.9\textwidth]{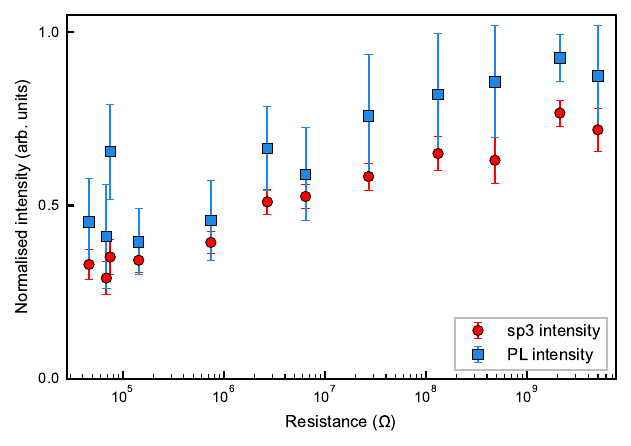}
\caption{\textbf{Broadband PL intensities analysis for wire electrodes.} Normalized broadband PL intensity (blue squares) and sp3 Raman intensity (red circles) plotted versus electrode resistance for laser-written wires at varying scan speeds. PL was collected for a spectral window of ~550-850~nm and both signals were normalized using high-sp3 ('pure-diamond') reference pixels (see Supporting Note~3). Error bars indicate 1$\sigma$.}\label{fig:SI_PL}
\end{figure}

To evaluate whether broadband photoluminescence (PL) intensity can serve as an in situ process indicator, we extracted the PL intensity from the written pixels. The results are presented in Supporting Figure \ref{fig:SI_PL}. This figure shows a correlation between the resistance and the normalized PL intensity of the written wire electrodes, which suggests that broadband PL intensity can be leveraged to assess performance of the written electrodes. 

Supporting Figure \ref{fig:SI_PL} also compares the PL with the sp3 intensity metric proposed in the main manuscript. This comparison showcases how the normalized PL follows the sp3 trend: PL increases with resistance, consistent with a larger fraction of diamond-like bonding. Yet, despite having similar correlation with the resistance of the electrodes, the PL normalized intensity presents significantly larger variance in the measurements. This is due to the broad spectral collection window of these PL maps ($\sim$550-850~nm), which can include emission from phenomena not directly related to graphitization (i.e., in Type~Ib diamond intrinsic defects generate a non-negligible background in PL maps). 

Thus, even if broadband PL can serve as a quantitative indicator during laser fabrication, in practice it is less specific and less accurate than the sp3 intensity, which should be favored for in situ process monitoring. 





\section*{Supporting Note 5: sp3 distribution}

\begin{figure}[H]
\centering
\includegraphics[width=1\textwidth]{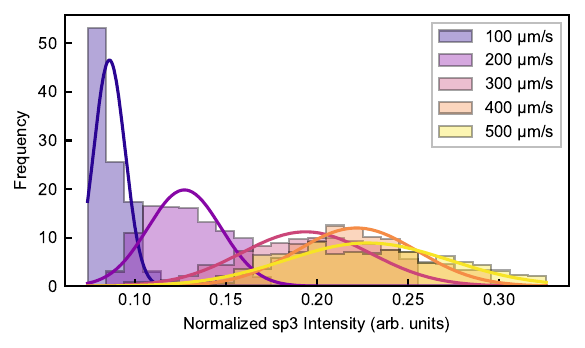}
\caption{\textbf{Distribution of normalized sp3 intensity for pad electrodes at different scan speeds.} Probability-density histograms with normal distribution fits of the normalized sp3 Raman intensity from pads written at different scan speeds (1500 spectra per laser scan speed). Increasing scan speed shifts the distribution to higher sp3 intensity and broadens it, indicating reduced graphitization and increased phase inhomogeneity.}\label{fig:S3_pad}
\end{figure}

\begin{figure}[H]
\centering
\includegraphics[width=1\textwidth]{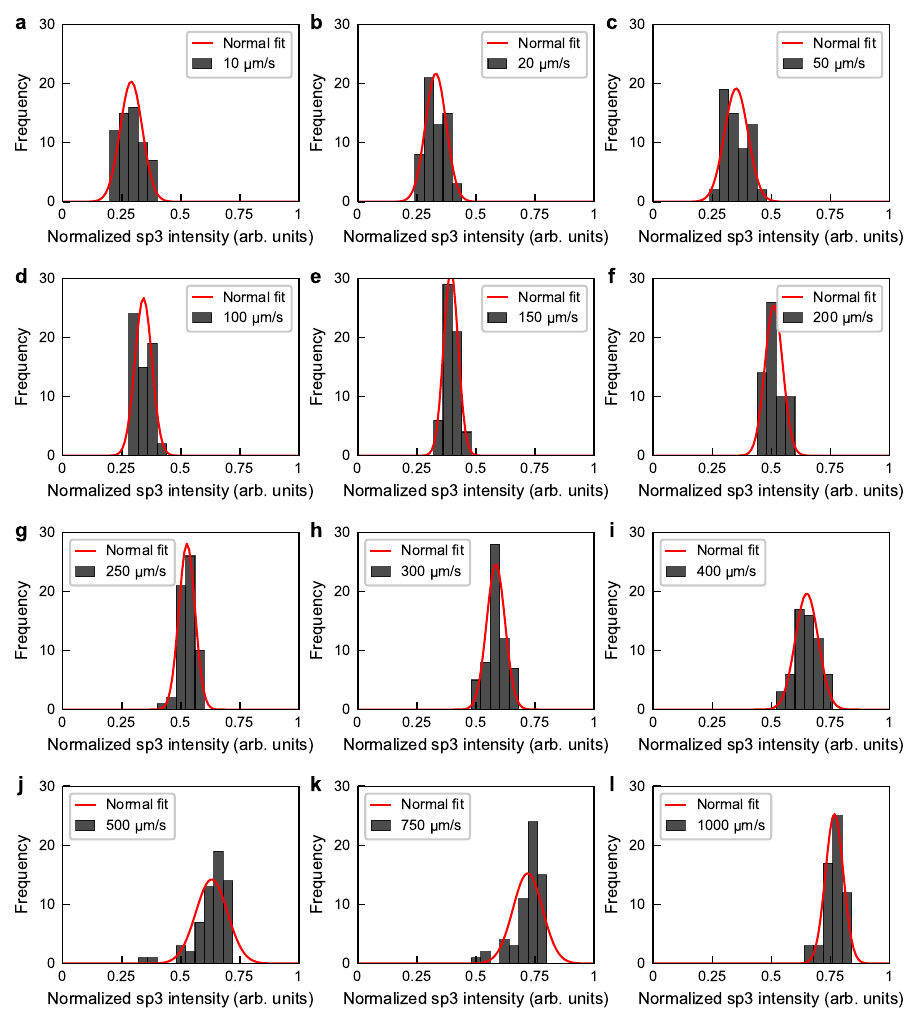}
\caption{\textbf{Distribution of normalized sp3 intensity for wire electrodes at different scan speeds.} Histograms of 60 pixels per electrode with corresponding normal distribution fits are shown for scan speeds ranging from 10~$\mu$m~s$^{-1}$ to 1000~$\mu$m~s$^{-1}$. The absence of overwriting in wire electrodes results in narrower distributions compared to pad electrodes. Increasing scan speed shifts the distributions toward higher sp3 values, consistent with reduced deposited energy per unit length and a lower degree of graphitization.}\label{fig:S3}
\end{figure}

The speed-dependent probability-density histograms of the normalized sp3 Raman intensity, computed from pixels identified as graphitized in the hyperspectral maps, are shown for pad and wire electrodes in Supporting Figure~\ref{fig:S3_pad} and Supporting Figure~\ref{fig:S3}, respectively. For each scan speed, 1500 pixels are analyzed for pads and 60 for wires. Intensities are referenced to pristine diamond, so higher normalized sp3 corresponds to weaker graphitization.

In pad electrodes, increasing scan speed shifts the distributions toward higher values and simultaneously broadens them, consistent with reduced energy deposition per unit length and a less uniform sp3-to-sp2 transformation. By contrast, wire electrodes exhibit narrower distributions across all scan speeds because their geometry prevents overwriting. Nonetheless, they display the same overall trend as pads: increasing scan speed shifts the distributions to higher normalized sp3 values, reflecting a lower degree of graphitization in the laser-written structures.

\section*{Supporting Note 6: sp3 Raman analysis}

\begin{figure}[H]
\centering
\includegraphics[width=0.7\textwidth]{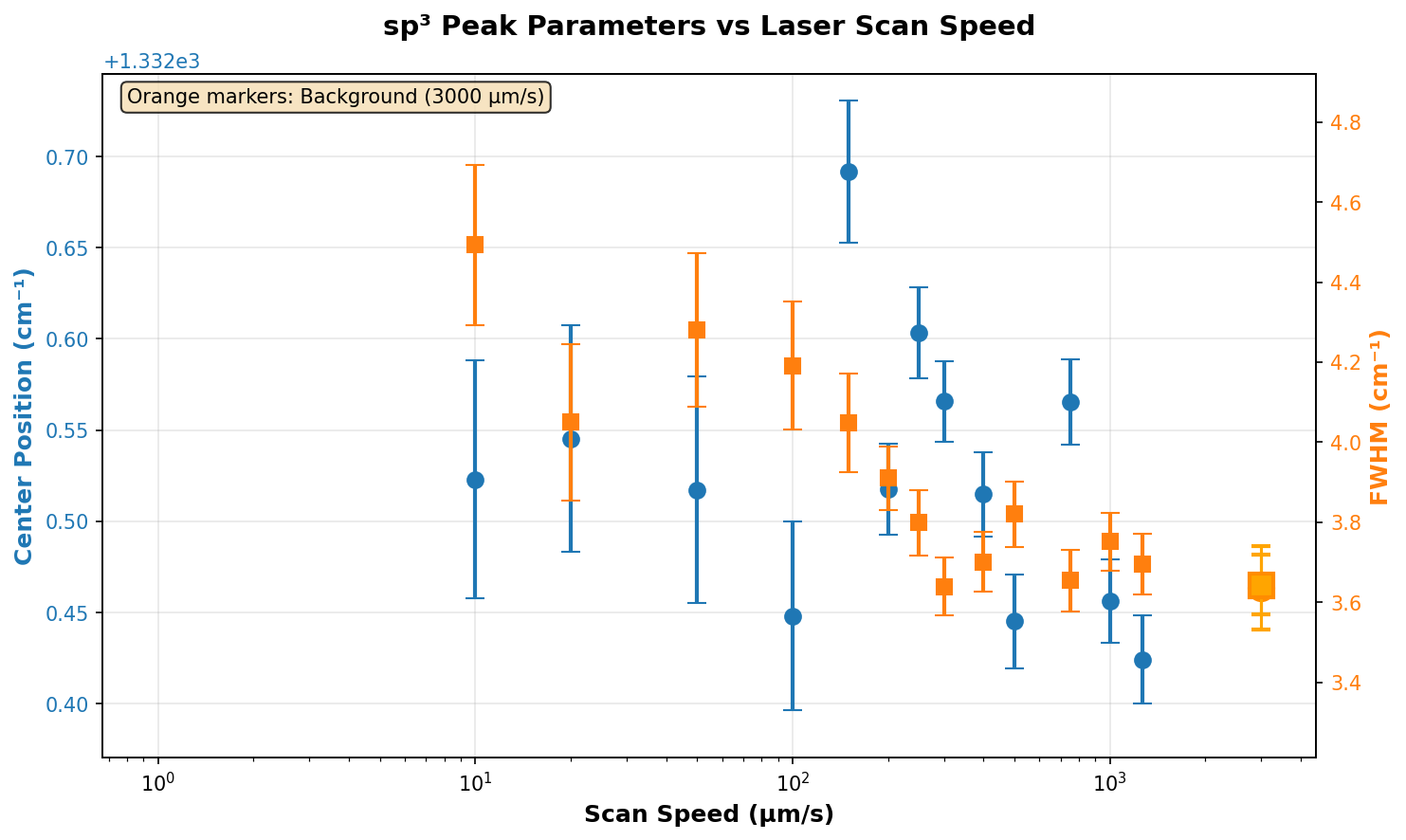}
\caption{\textbf{sp3 Raman peak evolution with laser scan speed in wire electrodes.} Mean center position (blue circles, left axis) and FWHM (dark-orange squares, right axis) extracted from Lorentzian fits to averages of 60 pixels within the graphitized wire region for each speed (10-1270~$\mu$m~s$^{-1}$). A pristine-diamond reference (100-pixel average from an unexposed area) is placed at 3000~$\mu$m~s$^{-1}$ for visual comparison. The center exhibits a slight blue shift relative to 1332~cm$^{-1}$, consistent with stress coupling between the graphitized layer and the surrounding diamond, while the FWHM narrows with increasing scan speed, indicating reduced disorder at higher speeds. Error bars indicate pixel-to-pixel variability.}\label{fig:S4a}
\end{figure}

To complement the two-probe transport data, we quantified how the sp3 Raman response varies with laser scan speed. For each speed, 60 pixels within the graphitized region of a laser-written wire were averaged and the sp3 peak was fitted with a Lorentzian. 
Averaged spectra for scan speeds from 10 to 1270~$\mu$m~s$^{-1}$ were analyzed; the resulting peak center and full width at half maximum (FWHM) are plotted in Fig.~\ref{fig:S4a} as blue circles and dark-orange squares, respectively. A pristine-diamond reference, obtained by averaging 100 pixels from an unexposed area, is shown at 3000$\mu$m~s$^{-1}$ for visual comparison only.

\begin{figure}[H]
\centering
\includegraphics[width=1\textwidth]{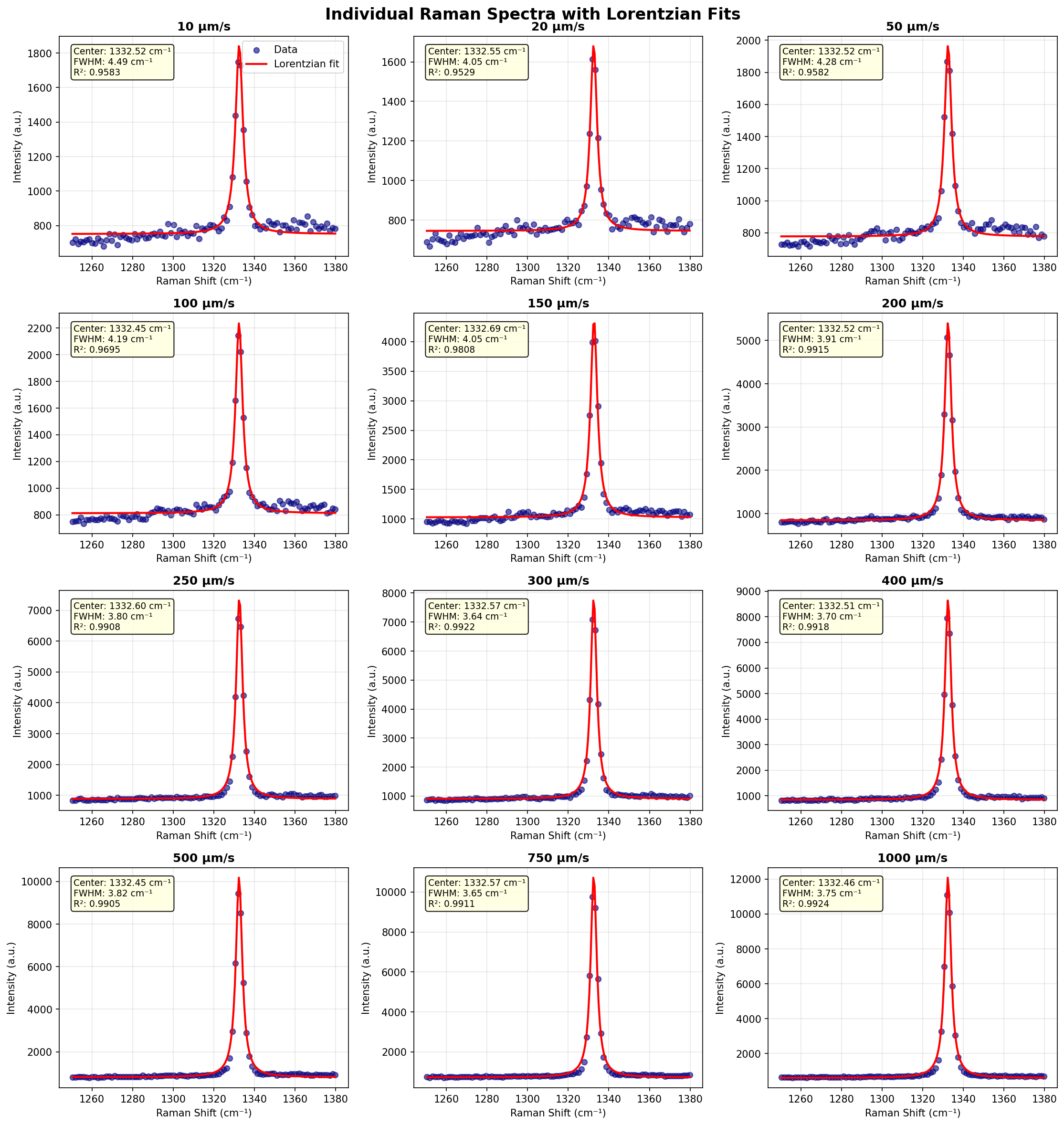}
\caption{\textbf{Representative sp3 Raman spectra and Lorentzian fits at different scan speeds.} Averaged spectra from 60 pixels within the graphitized wire region are shown for laser scan speeds ranging from 10 to 1000~$\mu$m~s$^{-1}$. Data are plotted as blue points and Lorentzian fits as red curves. The extracted peak center position, FWHM, and coefficient of determination ($R^2$) for each fit are indicated in the insets. At slower scan speeds, the sp3 mode broadens significantly, reflecting enhanced graphitization, lattice disorder, and strain distributions, while higher scan speeds produce narrower peaks, indicative of reduced disorder and improved crystalline retention.}\label{fig:S4b}
\end{figure}

Across the series, the sp3 mode lies slightly above 1332cm$^{-1}$, consistent with tensile stress in the transformed graphite layer and the concomitant in-plane compressive stress imparted to the adjacent diamond\cite{Wang2000}, which produces a blue shift of the diamond Raman mode~\cite{Grimsditch1975,Zaitsev2001}. The FWHM decreases systematically with increasing scan speed. Slower scans deposit more energy per unit length, driving a higher degree of graphitization; accordingly, the sp3 line departs from an ideal Lorentzian, broadening and sometimes becoming asymmetric. This behavior reflects increasing lattice disorder and strain distributions (inhomogeneous broadening), reduced phonon lifetimes (homogeneous broadening), and phonon-confinement effects; if the material becomes sufficiently conducting, Fano interference with an electronic continuum can further skew the lineshape~\cite{Orwa2000,Orwa2009}. All averaged spectra and their Lorentzian fits are provided in Fig.~\ref{fig:S4b}.

\section*{Supporting Note 7: Electrical performance for both pad and wire}

\begin{figure}[H]
\centering
\includegraphics[width=1\textwidth]{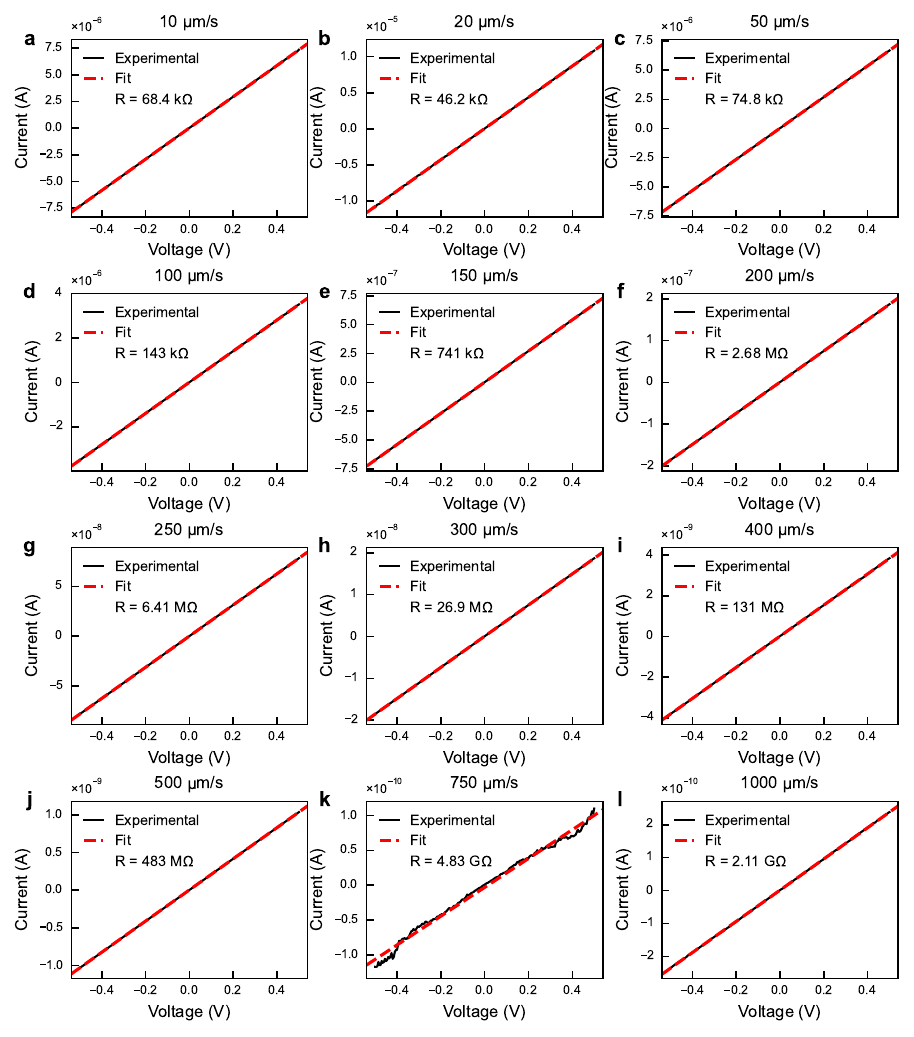}
\caption{\textbf{Scan-speed dependent transport in laser-written wire electrodes.}~\textbf{a-l}, Current-voltage characteristics (black) with linear fits (red, dashed) for wires written at 10, 20, 50, 100, 150, 200, 250, 300, 400, 500, 750, and 1000~$\mu$m~s$^{-1}$, measured over $-0.5$ to $+0.5$~V. All traces are ohmic and the fitted resistances increase by orders of magnitude as the scan speed rises, evolving from the k$\Omega$ regime at slow scans to the M$\Omega$-G$\Omega$ range at fast scans. The reduction in resistance at slow scans correlates with the depletion of sp3 Raman intensity and the concurrent strengthening of sp2 signatures, evidencing progressive graphitization.}
\label{fig:S5}
\end{figure}

Electrical measurements confirm that femtosecond-laser writing produces ohmic graphitic pathways whose conductance scales with scan speed and tracks the depletion of sp3 Raman intensity in the written region. For wire electrodes, the I-V characteristics are strictly linear from -0.5 to +0.5~V, and linear fits yield resistances that vary over several orders of magnitude as the scan speed increases from 10 to 1000~$\mu$m~s$^{-1}$ (Fig.~\ref{fig:S5}a-l). At the slowest scans (10-50~$\mu$m~s$^{-1}$) the wires exhibit resistances in the tens of k$\Omega$, approaching the series resistance floor of the test board contact pads, whereas progressively faster scans—depositing less energy per unit length—produce markedly higher resistances reaching the M$\Omega$-G$\Omega$ range.

\begin{figure}[H]
\centering
\includegraphics[width=1\textwidth]{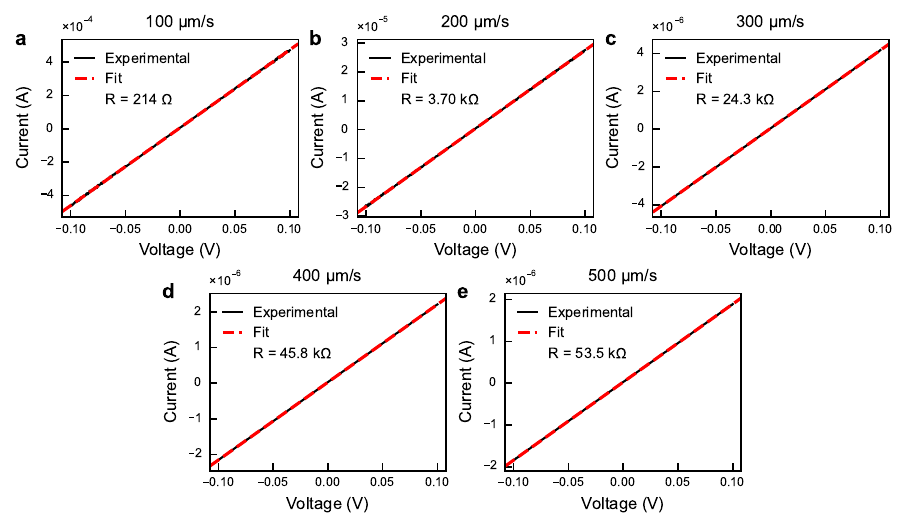}
\caption{\textbf{Pad electrodes: lower absolute resistance with the same scan-speed trend.}~\textbf{a-e)} IV characteristics (black) and linear fits (red, dashed) for pads written at 100, 200, 300, 400, and 500~$\mu$m~s$^{-1}$, measured over $-0.1$ to $+0.1$~V. The response is strictly ohmic, with fitted resistances rising from $\sim$214~$\Omega$ at 100~$\mu$m~s$^{-1}$ to $\sim$53.5~k$\Omega$ at 500~$\mu$m~s$^{-1}$. Compared with wires, pads exhibit substantially lower absolute resistance due to their larger cross-section and multi-track overlap. As in the wires, the decrease in resistance at slower scans tracks the depletion of sp3 intensity and the growth of sp2 features in the Raman spectra, confirming a direct Raman-electrical correlation.}

\label{fig:S5_pad}
\end{figure}

Pad electrodes follow the same trend but with substantially lower absolute resistance due to their larger cross-section and multi-track overwriting, which yields more complete graphitization as indicated by the depleted sp3 Raman signal. I-V curves acquired from -0.1 to +0.1~V remain perfectly linear, with fitted resistances rising from a few hundred $\Omega$ at 100~$\mu$m~s$^{-1}$ to $\sim$50k~$\Omega$ at 500~$\mu$m~s$^{-1}$ (Fig.~\ref{fig:S5_pad}a-e). The comparison between wires and pads highlights the expected geometric scaling and the writing process: wider, thicker pads support significantly higher conductance than narrow wires under otherwise identical conditions, and the zig-zag writing strategy introduces controlled overwriting that further enhances graphitization.

The electrical evolution with scan speed mirrors the Raman signatures of the laser-induced carbon. In both geometries, depletion of sp3 intensity and the concurrent strengthening of sp2 features at slower scan speeds coincide with reduced resistance, whereas the relative persistence of sp3 character at faster scans coincides with increased resistance. This direct Raman-electrical correlation substantiates that efficient graphitization—conversion from sp3 to sp2 bonding—governs charge transport in the written structures.

\section*{Supporting Note 8: Hyperspectral unmixing}

\begin{figure}[H]
\centering

\includegraphics[width=\textwidth]{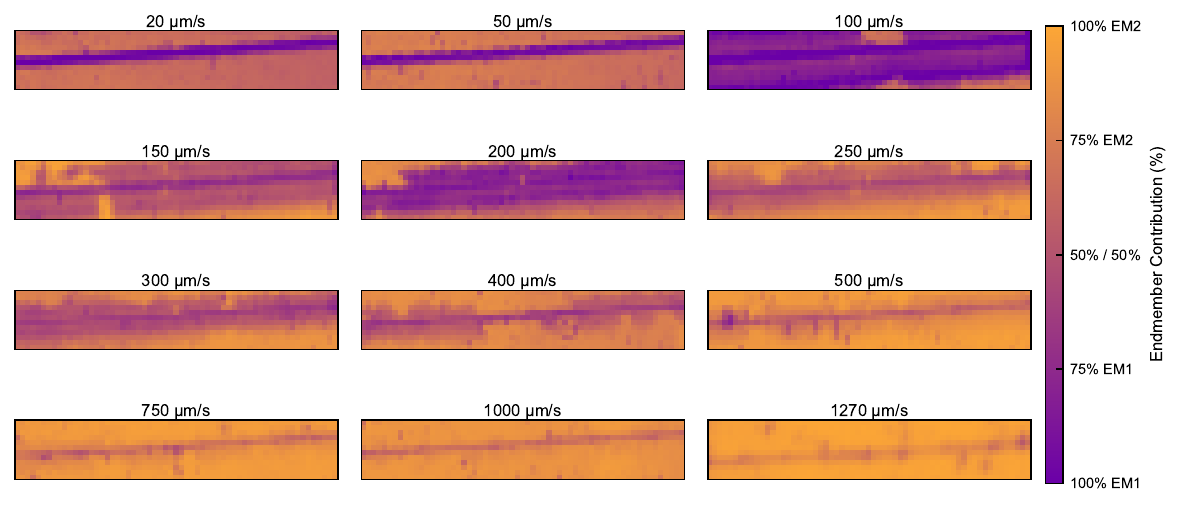}\\
\textbf{(a) Spectral unmixing (VCA) on all wires.}

\vspace{1em} 

\includegraphics[width=\textwidth]{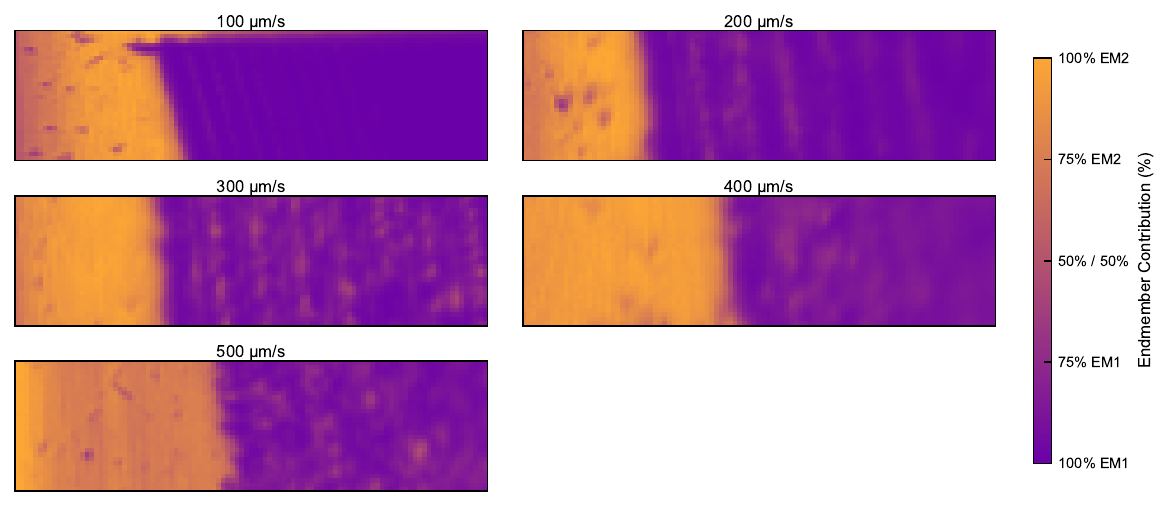}\\
\textbf{(b) Spectral unmixing (VCA) on all pads.}

\caption{\textbf{Spectral unmixing (VCA) on all wires and pads.} These abundances maps are obtained applying vertex component analysis unmixing with fully-constraint linear least squares endmember determination to each complete dataset independently. The specific endmembers can be found in Supporting Figure \ref{fig:SI_unmix}.}
\label{fig:SI_unmix}
\end{figure}

We performed hyperspectral unmixing with a linear mixing model using vertex component analysis (VCA)~\cite{nascimento2005vertex} for endmember extraction and fully-constrained least squares (FCLS) for abundance estimation. As shown in Supporting Figure~\ref{fig:SI_unmix}, the unmixing results consistently show progressive loss of the diamond-associated component with decreasing scan speed, while the carbon-associated component increases and eventually saturates. This reinforces the main conclusion that sp$^{3}$ depletion provides a reliable, monotonic indicator of graphitization. Hyperspectral unmixing is a powerful technique that leverages the full spectrum and does not require pre-defined labels. However, it is sensitive to the presence of noise or species not considered in the analysis. As shown in Supporting Figure \ref{fig:SI_unmix}a, the presence of the broadband emission damage shell when fabricating at intermediate speeds is mislabeled as being a mixture of diamond and graphite phases, since in the analysis only two endmembers are considered.

\begin{figure}[H]
\centering

\includegraphics[width=\textwidth]{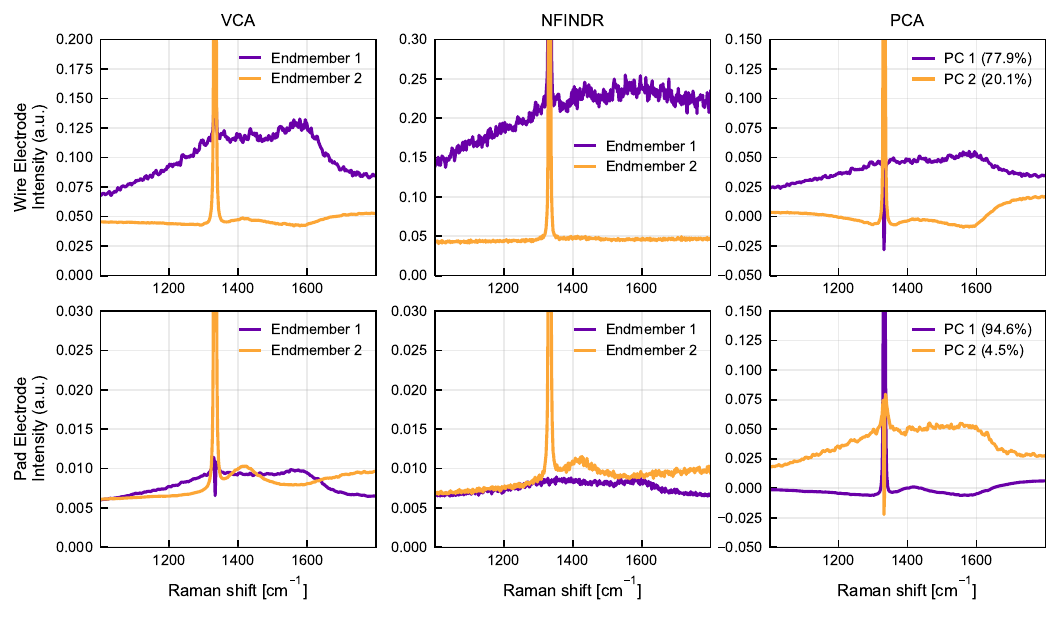}\\
\textbf{(a) Endmembers when unmixing in the spectral ROI including sp3 (1000-1800~cm$^{-1}$).}

\vspace{1em} 

\includegraphics[width=\textwidth]{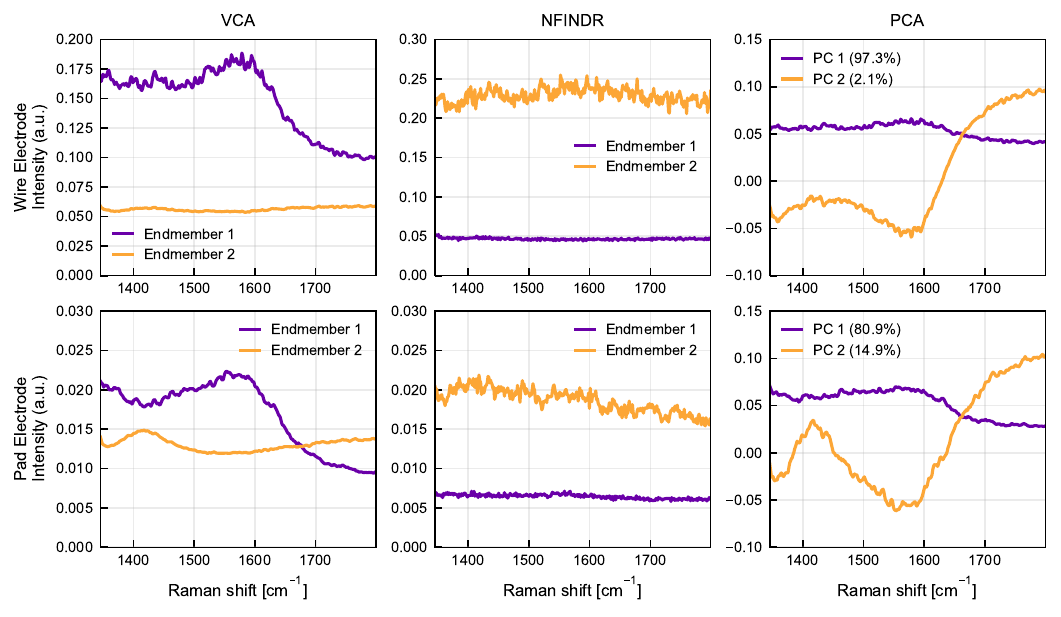}\\
\textbf{(b) Endmembers when unmixing in the spectral ROI not including sp3 (1345-1800~cm$^{-1}$).}

\caption{\textbf{Endmembers for pad and wire electrodes.} Endmembers for the wire and pad electrodes using different unmixing methods. Also, the first two principal components resulting from principal component analysis, with their respective explained variance.}
\label{fig:SI_unmixing_endmembers}
\end{figure}

Supporting Figure~\ref{fig:SI_unmixing_endmembers} presents the endmembers extracted using different unmixing techniques and spectral regions of interest (ROI), exhibiting the variance that these endmembers can have. This behavior is expected: VCA and \textit{N-FINDR}~\cite{winter1999n} are geometric algorithms that approximate the data as lying within a convex simplex. They perform best when pixels lie close to the true extremes, but in practice the confocal voxel contains mixtures, variable SNR, NV$^{0}$ fluorescence, and potentially stress-induced peak shifts. The algorithms therefore identify different “extreme’’ representatives of the diamond-like and graphite-like responses, leading to small method-dependent variations that do not alter the physical interpretation. Indeed, the endmembers preserve the spectral features of their associated spectra, that is, the sp3 peak and NV0 band for the endmember associated to diamond, and the G band and, to lesser extent, the D band for the one associated to graphite.

Notably, for the spectral ROI including the sp3 peak, the large dynamic range of this feature amplifies small fluctuations which can distort the convex geometry and bias endmember selection. Excluding this line (spectral ROI: 1345–1800~cm$^{-1}$) reduces leverage and produces noisier but qualitatively similar maps.

Supporting Figure~\ref{fig:SI_unmixing_endmembers} also provides the first two principal components.
It is important to distinguish geometric unmixing from principal component analysis (PCA)~\cite{abdi2010principal}. VCA and N-FINDR seek spectra at the boundaries of the data cloud that can be interpreted as approximate endmembers, whereas PCA decomposes variance into orthogonal components that need not resemble real spectra and can take negative values. PCA is valuable for denoising and diagnostics, but its components cannot be interpreted as abundances.

\newpage
\bibliographystyle{unsrt} 
\bibliography{Lit}

\end{document}